\documentclass[aip,jcp,amsmath,amssymb,floatfix,letterpaper,groupedaddress,preprint]{revtex4-2} 
\usepackage[margin=1in]{geometry}

\usepackage{physics} 
\usepackage{mhchem} 
\usepackage{siunitx}
\usepackage{amsthm} 
\usepackage{pgfplots} 

\usepackage{subfloat} 
\usepackage{float} 
\usetikzlibrary{arrows.meta} 
\usepackage{braket} 
\usepackage{subfig}

\usepackage{amsmath}
\usepackage{amssymb}
\usepackage{mathtools} 

\usepackage{import} 
\usepackage{lettrine} 
\usepackage{array}

\usepackage{bm} 
\usepackage{multirow} 
\usepackage{IEEEtrantools} 

\usepackage{enumitem}               

 \usepackage{nicefrac}               

\newcommand{\Sref}[1]{Section~\ref{#1}}
\newcommand{\Aref}[1]{Appendix~\ref{#1}}
\newcommand{\Eref}[1]{Eq.~(\ref{#1})}
\newcommand{\Fref}[1]{Fig.~\ref{#1}}

\newlength{\adjustment}
\newlength{\figurewidth}
\newlength{\eqspace}
\usepackage[title]{appendix}

\begin{document}

\title{Normal ordered exponential approach to thermal properties and time-correlation functions: General theory and simple examples}
\author{Marcel Nooijen}
\affiliation{Department of Chemistry, University of Waterloo, Waterloo, Ontario, Canada, N2L 3G1}
\email{nooijen@uwaterloo.ca}
\author{Songhao Bao}
\affiliation{Department of Chemistry, University of Waterloo, Waterloo, Ontario, Canada, N2L 3G1}
\date{\today}

\begin{abstract}
A normal ordered exponential parametrization is used to obtain equations for thermal one-and two-particle reduced density matrices, as well as free energies, partition functions and entropy for both Fermionic (electronic) and Bosonic (vibrational) Hamiltonians. A first principles derivation of the equations, relying only on a simple Wick's theorem and starting from the differential equation $\frac{d \hat{D}}{d \beta}= - (\hat{H}-\mu \hat{N})\hat{D}$, is presented that yields a differential equation for the amplitudes representing density cumulants, as well as the grand potential. In contrast to other approaches reported in the literature the theory does not use perturbation theory in the interaction picture and an integral formulation as a starting point, but rather requires a propagation of the resulting differential equation for the amplitudes. While the theory is applicable to general classes of many-body problems in principle, here, the theory is illustrated using simple model systems. For one-body Fermionic Hamiltonians, Fermi-Dirac one-body reduced density matrices are recovered for the grand-canonical formulation. For multidimensional harmonic oscillators numerically exact results are obtained using the thermal normal ordered exponential (TNOE) approach. As an application of the related time-dependent formulation numerically exact time-autocorrelation functions and absorption spectra are obtained for harmonic Franck Condon problems. These examples illustrate the basic soundness of the scheme and are used for pedagogical purposes. Other approaches in the literature are only discussed briefly and no detailed comparative discussion is attempted.
\end{abstract}
\maketitle

\section{\label{sec:Introduction}Introduction}
The calculation of thermal properties is important for many branches of chemistry and physics. The easiest realization perhaps is in thermochemistry of (ideal) gases, in which \emph{nuclear motion} is commonly treated through free particle, rigid rotor and harmonic approximations. The calculation of thermal properties due to nuclear motion in liquids or flexible proteins is most often treated using sampling of classical motion. 
Calculations of thermal \emph{electronic} properties have a long history in the context of solid state physics, in particular for gap-less materials, e.g. metals, and/or small gap materials, for example in Hubbard-like models for strongly correlated materials. Hartree-Fock or more commonly density functional theory is widely applied in the convenient grand-canonical formulation in which one obtains fractional occupation numbers according to the Fermi-Dirac distribution, $n_p = (1+ e^{\beta( \epsilon_p - \mu)}) ^{-1}$. The chemical potential in such calculations is adjusted to achieve  electrical neutrality.  The other most widely applied strategies to obtain thermal properties for solids use a variety of approximations rooted in many-body Green's function theory, e.g. \cite{fetter2012quantum,abrikosov2012methods,negele2018quantum,mattuck1992guide}. Such approaches apply to both Fermionic (i.e. electronic) and Bosonic (e.g. vibrational) degrees of freedom and in this paper we will likewise explore both types of problems from a unified perspective.

 Thermal electronic properties are also of interest for molecular systems, e.g. gases or liquids that contain low-lying electronic states, and recent years have witnessed an increased interest in the calculation of thermal properties using first principles theories, e.g. through perturbation theory \cite{matsubara1955new,bloch1958developpement, hirata2013kohn,hirata2013second,he2014finite,santra2017finite}, second-order Green's function approaches, e.g. \cite{welden2016exploring,kananenka2016efficient,kananenka2016efficientb,zgid2017finite},  or variations of thermal coupled cluster theory \cite{sanyal1992thermal,mandal1998thermal,mandal2001non,mandal2003finite,hermes2015finite,
white2018time-dependent,hummel2018finite}. While Coupled Cluster theory today is arguably the most versatile and accurate wave function based theory for electronic ground states, the theoretical foundation of the various Coupled Cluster formulations to obtain thermal properties are somewhat curious, and perhaps ad hoc. Most of the formulations are justified according to the following principles:
\begin{itemize}
\item [(a)] At zero temperature the thermal coupled cluster approach reduces to the conventional single reference coupled cluster formulation for the ground state.
\item[(b)] A low-order perturbation expansion of the thermal CC approach agrees with established finite temperature perturbation theory (e.g. in the Matsubara formulation).
\item[(c)] The formulations explicitly use the interaction picture to derive equations, and involve an integration over inverse temperature.
\item[(d)] Thermal energies and other thermal properties are obtained as a function of the inverse temperature $\beta$ and the chemical potential $\mu$, and are not strictly limited to systems with a specified number of electrons in their ensemble average.  
\end{itemize}
It would appear that this state of affairs has mainly historical origins. Much of the original work on thermal properties has been done based on a quantum field theoretical framework, extending time-dependent propagator or Green's function theory to thermal properties, using in essence a Wick rotation, in addition to further analysis. In this historical context a perturbative diagrammatic starting point is the essential point of departure, and any resulting theory is naturally rooted in perturbative expansions and partial re-summations. The PhD work of one of the authors took exactly this point of view, starting from perturbative diagrammatic expansions, and back-engineering to infinite order theories like Coupled Cluster theory and the Coupled Cluster Green's function, using recursive diagrammatic procedures \cite{nooijen1992thesis,nooijen1992coupled,nooijen1993coupled}. Interestingly, in this process the Green's function was obtained directly, and not through an approximation to the irreducible self-energy and a subsequent solution of Dyson's equation. This  suggests that the methods of quantum field theory, which are defined from the outset in terms of a perturbation expansion may benefit from a reformulation when a closed form Hamiltonian is known. In the context of quantum chemistry, the picture is usually reversed. One has easy direct ways to define Coupled Cluster theory, as an approximation to full CI, and a perturbative expansion of the theory leads to many-body perturbation theory, in a manifestly connected form, e.g. \cite{monkhorst1981recursive,harris1992algebraic,shavitt2009many}.

The tight connection to essentially single reference perturbation theory is likely not desirable if the goal is to calculate thermal properties for molecular systems, including explicitly electronic degrees of freedom. Molecular systems of interest will have multiple low-lying electronic states, otherwise one need not bother with electronic contributions to thermal properties beyond the ground state. The most likely realistic systems would require a multireference description for their ground states, and would be poorly described by (low-rank) single reference coupled cluster theory or many-body perturbation theory. The logical conclusion would be that if one is interested in efficient methodologies to calculate thermal properties (of strongly correlated systems),  methods should \emph{not} reduce to conventional single reference Coupled Cluster theory at zero temperature, and/or one would prefer low-order perturbative schemes that do \emph{not} reduce to conventional finite temperature perturbation theory, as these methods do not work well for strongly correlated systems. We think this is a fair assessment for molecular systems, although the situation for metallic solids may be different. 

We think the above arguments strongly suggest one should take an alternative tack, avoiding the conventional routes. Rather than starting from a complicated framework rooted in Quantum Field Theory the basic starting point for molecular calculations can be the full CI formulation in a finite basis set, using either canonical or grand-canonical formulations. For small, but meaningful, strongly correlated systems full CI thermal properties are readily obtained, and the goal of more efficient, approximate formulations would be to approach the full CI results, implying one needs to design a formulation that can work with low-rank, e.g. singles and doubles substitution operators. There does not seem to be any pressing need to use the interaction picture, as all such approaches are rooted in perturbation theory that would seem to be a poor starting point. It may be that newly derived approaches could be reformulated using the interaction picture, but we do not want to make this a point of departure.

This paper is largely pedagogical in nature. We present a general theory to calculate thermal properties for both Fermionic and Bosonic systems and will also consider closely related time-dependent formulations. The approach is based on a normal ordered exponential ansatz to represent the many-body density matrix and the use of a simple Wick's theorem (i.e. no interaction picture) to derive ordinary coupled differential equations (in terms of a single parameter $\tau$) for the amplitudes that can simply be solved for, given suitable initial conditions. The goal of the paper is to clearly explain the theory from simple principles, and we will use simple examples like one-electron Hamiltonians and displaced harmonic oscillators for which exact (analytical) solutions are known. As an example of a time-dependent property we will discuss the calculation of time-autocorrelation functions and related absorption spectra for general harmonic Franck-Condon problems. The implementation of these methods for more general problems is of course of real interest, but a number of issues are expected to arise and here we focus on simple problems for which the proposed methods are numerically exact.

This paper is organized as follows. In \Sref{sec:elec-general} we discuss the general formulation of the electronic problem for the grand-canonical ensemble. In \Sref{sec:elec-one} we apply the Fermionic theory to one-electron hamiltonians and discuss how the results reduce to conventional Fermi-Dirac results. In \Sref{sec:vib-general} we discuss the general bosonic case for (single surface) vibrational Hamiltonians. We illustrate the theory for simple 1d harmonic oscillators in \Sref{sec:vib-1d} and recover Bose-Einstein statistics.  In the appendix we provide the formulation for  general multidimensional haromic hamiltonian and this can be generalized to more complicated Hamiltonians. In \Sref{sec:FC-spectra} we show how the theory can be adjusted to time-dependent problems and we show how it provides a convenient way to calculate Harmonic Franck-Condon spectra, with an application to the photo-electron spectrum of formaldehyde. In \Sref{sec:reflections} we have another pedagogical look at the three types of statistics (Fermi-Dirac, Bose-Einstein and Boltzmann) and discuss how they are related by 'slightly' different differential equations. We end the paper with some concluding and summarizing remarks and a further outlook in \Sref{sec:conclusion}.

\section{\label{sec:elec-general} Thermal Normal Ordered Exponentials for Fermions and the grand canonical ensemble}
The starting point for the discussion is the differential equation for the thermal density matrix in the grand canonical formulation
\begin{IEEEeqnarray}{rLl}
- \frac{d\hat{D}}{d{\beta}}&=( \hat{H}- \mu \hat{N}) \hat{D} \\
\hat{D}(\beta=0) &= \hat{D}_0= \alpha \hat{1}
\end{IEEEeqnarray}
 We define a uniform fermionic density operator, $\hat{D}_0$ at $\beta=0$, and this serves as the vacuum for the many-body theory. In the definition above $\alpha$ is a (formal) normalization constant such that $Tr(\hat{D}_0)=1$. To derive equations we trace over the complete Fock space using the uniform density, but using Wick's theorem we only require the notion of normal order and contractions between elementary annihilation and creation operators. We define the non-zero contractions as 
\begin{IEEEeqnarray}{rLl}
\hat{p}^\dagger \cdot \hat{q} &= \Trace \left( \hat{p}^\dagger \hat{q} \hat{D}_0 \right) \equiv \left\langle \hat{p}^\dagger \hat{q} \right\rangle \equiv f_p \delta_{pq} = f \delta_{pq} \\
\hat{p} \cdot \hat{q}^\dagger &= \delta_{pq}- \left\langle \hat{q}^\dagger \hat{p} \right\rangle =(1-f_q)\delta_{pq} \equiv  \bar{f}_q \delta_{pq} = \bar{f} \delta_{pq}
\end{IEEEeqnarray}
Due to the uniformity of the density operator the value of the single parameter $f$ is determined by the condition that the trace of the one-body density matrix should equal the number of electrons, $n_{el}$. If we have $M$ orbitals in total in the orthonormal one-particle basis set we have
\begin{IEEEeqnarray}{rLl}
\sum_p f &= n_{el};  \qquad   f = \frac{n_{el}}{M}, \bar{f} = \frac{M-n_{el}}{M}
\end{IEEEeqnarray}

The theory we will develop is completely based on a simple form of Wick's theorem where the product of two normal operators is defined as their normal product with all possible contractions. Another important property is that the vacuum expectation value of any normal-ordered operator vanishes except for the constant part
\begin{IEEEeqnarray}{rLl}
Tr (\left\lbrace \hat{\Omega}_{\lambda} \right\rbrace \hat{D}_0) \equiv \braket {\left\lbrace  \hat{\Omega}_{\lambda} \right\rbrace} = 0
\end{IEEEeqnarray}
for any nonempty string of annihilation and creation operators $\hat{\Omega}_{\lambda}$.
In this work we can use a very familiar version of Wick's theorem, where the only generalization is that all orbitals are treated equivalently and there is no distinction between occupied and virtual orbitals.

The finite temperature many-body density operator is parameterized using a normal ordered exponential as 
\begin{IEEEeqnarray}{rLl}
\hat{D} = \left\lbrace e^{\hat{S}(\beta)}\right\rbrace \hat{D}_0
\end{IEEEeqnarray}
where the braces indicate normal ordering and $\hat{S}(\beta)$ is expanded in terms of normal ordered substitution operators and a constant term
\begin{IEEEeqnarray}{rLl}
\hat{S}(\beta) &= s_0(\beta) + \sum_{\lambda} s_{\lambda}(\beta)\left\lbrace \hat{\Omega}_{\lambda} \right\rbrace 
\end{IEEEeqnarray}
The expansion can be truncated after some highest rank of substitution. The complete untruncated operator may require an expansion up to the dimension of the Fock space. To work with the normal ordered exponential one can simply use the Taylor series expansion for the exponential and use that there are no contractions between $\hat{S}$ operators when evaluating Wick's theorem, e.g. 
\begin{IEEEeqnarray}{rLl}
 \left\lbrace e^{\hat{S}(\beta))}\right\rbrace = \sum_{n=0} \frac{1}{n!} \left\lbrace (\hat{S}(\beta))^n\right\rbrace
\end{IEEEeqnarray}
To facilitate the discussion we partition normal ordered operators in a constant part and the remaining part of the operator, which has non-trivial (normal ordered) substitution operators. This non-trivial part of the operator will be denoted through a \emph{dot} rather than a \emph{carrot}, e.g. 
\begin{IEEEeqnarray}{rLl}
\hat{S}(\beta) = s_0(\beta) + \dot{S}(\beta) \\
\hat{N} = n_{el} + \dot{N} \\
\hat{H} = E_0 + \dot{H} 
\end{IEEEeqnarray}
Due to the normal ordering convention the trace of the density matrix is easily evaluated as
\begin{IEEEeqnarray}{rLl}
Tr(\hat{D}) &= Tr\left\lbrace exp(s_0(\beta) + \dot{S}(\beta)) \right\rbrace \hat{D}_0 \nonumber \\
&= e^{s_0(\beta)}  \left\langle  \left\lbrace  exp(\dot{S}(\beta)) \right\rbrace \right\rangle \nonumber \\
&= e^{s_0(\beta)}
\end{IEEEeqnarray}
To evaluate the thermal expectation value of any operator $\hat{O} = O_0 + \dot{O}$ one evaluates
\begin{IEEEeqnarray}{rLl}
\left\langle O \right\rangle &= \frac{Tr(\hat{O} \hat{D})}{Tr (\hat{D})} \nonumber \\
&= \left\langle (O_0 + \dot{O}) \left\lbrace e^{s_0(\beta)}exp(\dot{S}(\beta))\right\rbrace \right\rangle e^{-s_0(\beta)}
&= O_0 + \left\langle  \dot{O} \left\lbrace exp(\dot{S}(\beta)) \right\rbrace \right\rangle 
\end{IEEEeqnarray}
where we note that the partition function $Z=e^{s_0}$ cancels between numerator and denominator. This is convenient as the partition function depends on an arbitrary choice for zero energy and may diverge for low T, if that zero of energy is above the ground state energy of the system.
It follows that if one knows the amplitudes of the operator $\dot{S}(\beta)$ one can evaluate expectation values over the thermal density matrix, while 
\begin{IEEEeqnarray}{rLl}
s_0(\beta) = ln (Tr(\hat{D})) &= ln Tr (e^{-\beta (\hat{H} - \mu \hat{N})}) = ln(Z(\beta,\mu))
\end{IEEEeqnarray}
is the grand potential, relating to the (Helmholtz) free energy, $A=-k_B T ln(Z) = -\frac{1}{\beta} ln (Z)$

 As we will see the constant term, $E_0$, in the hamiltonian is easily incorporated, like a shift in energy scale, and we will use as a starting equation for the $S$ amplitudes
\begin{IEEEeqnarray}{rLl}
- \frac{d\hat{D}}{d{\beta}}&=( \dot{H}- \mu \hat{N}) \hat{D} 
\end{IEEEeqnarray}
Substituting the normal ordered exponential parameterization in the differential equation for the density matrix, 
\begin{IEEEeqnarray}{rLl}
- \frac{d}{d{\beta}} \left\lbrace exp(\hat{S}(\beta)) \right\rbrace \hat{D}_0 &=( \dot{H}- \mu \hat{N}) \left\lbrace exp(\hat{S}(\beta)) \right\rbrace \hat{D}_0 
\end{IEEEeqnarray}
we can simplify immediately, using Wick's theorem (see appendix \Aref{sec:wick})
\begin{IEEEeqnarray}{rLl}
-  \left\lbrace \frac{d\hat{S}}{d\beta} exp(\hat{S}(\beta)) \right\rbrace \hat{D}_0 &= \left\lbrace \left[ ( \dot{H}- \mu \hat{N}) \left\lbrace exp(\dot{S}(\beta)) \right\rbrace \right]_{connected} exp(\hat{S}(\beta)) \right\rbrace  \hat{D}_0 
\end{IEEEeqnarray}
which implies the connected form of the equation
\begin{IEEEeqnarray}{rLl}
-   \frac{d\hat{S}}{d\beta} \hat{D}_0 &=  \left[ ( \dot{H}- \mu \hat{N}) \left\lbrace exp(\dot{S}(\beta)) \right\rbrace  \right]_{connected}   \hat{D}_0 
\end{IEEEeqnarray} 
The subscript ``connected" implies the string of operators is written in normal order, and one keeps only connected terms. The usual combinatorics of the exponential regenerates a normal ordered exponential as indicated above. We can equate the connected parts on both sides of the equation and project against a complete set of normal ordered substitution operators commensurate with the definition of maximum substitution rank to get a differential equation for the amplitudes
\begin{IEEEeqnarray}{rLl} \label{eq_grand_amp}
- \left\langle   \dot{\Omega}_{\nu}  \frac{d\dot{S}}{d\beta} \right\rangle &= \left\langle  \dot{\Omega}_{\nu}   \left[ ( \dot{H}- \mu \dot{N}) \left\lbrace exp(\dot{S}(\beta)) \right\rbrace \right]_{connected} \right\rangle 
\end{IEEEeqnarray}
\begin{IEEEeqnarray}{rLl} \label{eq_grand_s0}
-  \frac{ds_0}{d\beta} &= \left\langle   \left[ ( \dot{H}- \mu \hat{N}) \left\lbrace exp(\dot{S}(\beta)) \right\rbrace \right]_{connected} \right\rangle 
\end{IEEEeqnarray}
In our replacement of $\hat{N}$ by $\dot{N}$ in the amplitude equations we have used that a constant term cannot be connected to, or contracted against, other operators.

The number of electrons is determined as 
\begin{IEEEeqnarray}{rLl} 
N_{el} &= n_{el} + \left\langle  \left[ \dot{N} \left\lbrace exp(\dot{S}(\beta)) \right\rbrace \right]_{connected} \right\rangle
\end{IEEEeqnarray}
and the parameter $\mu(\beta)$ is to be determined such that $N_{el}$ evaluates to $n_{el}$ for all $\beta$. This implies the convenient relation
\begin{IEEEeqnarray}{rLl} \label{constr_N_el}
 \left\langle  \left[ \dot{N} \left\lbrace exp(\dot{S}(\beta)) \right\rbrace \right]_{connected} \right\rangle &=0
\end{IEEEeqnarray}
The equations \ref{eq_grand_amp}, \ref{eq_grand_s0}, \ref{constr_N_el} have to be solved in a coupled fashion. We note that $s_0(\beta)$, which is an extensive quantitity, does not enter the amplitude equations \ref{eq_grand_amp}, which is to be expected if a theory is to scale correctly with the size of the system. The initial condition for the integration over $\beta$ is simply $s_{\lambda}(\beta=0)=0 \quad \forall  \lambda$, starting from the high (or infinite) T limit $\beta=0$.

The thermal internal energy is given by 
\begin{IEEEeqnarray}{rLl}
U(\beta) &=E_0 + \left\langle  \left[ \dot{H} \left\lbrace exp(\dot{S}(\beta)) \right\rbrace \right]_{connected} \right\rangle 
\end{IEEEeqnarray}
The grand canonical partition function then satisfies the differential equation
\begin{IEEEeqnarray}{rLl}
- \frac{ln(Z)}{d\beta} &= E_0  -  \frac{ds_0}{d\beta} 
\end{IEEEeqnarray}
By integrating $\frac{ds_0}{d\beta}$ we obtain (the ln of) the grand canonical partition function
\begin{IEEEeqnarray}{rLl}
ln(Z(\beta)) &= - \beta E_0  + s_0(\beta)
\end{IEEEeqnarray}
Identifying temperature $T=(k_B \beta)^{-1}$, we can make the connection to thermodynamic electronic properties in the grand canonical ensemble through
\begin{IEEEeqnarray}{rLl}
- k_B T \hspace{\eqspace} ln (Z(T)) = U(T) - T \hspace{\eqspace} S(T) + \mu(T) \hspace{\eqspace} n_{el}
\end{IEEEeqnarray}
It follows that entropy can be obtained from the above formulation. The above relations will be verified numerically for a simple one-electron model problem in \Sref{sec:elec-one}, in which we can explicitly perform a sum over states expression and use elementary methods to obtain all quantities. For one-electron porblems we are recovering the correct results known from Fermi-Dirac statistics.

The one- and two-body reduced thermal density matrices can be obtained in a similar fashion as the internal energy, and only require the one- and two-body operators in $\dot{S}(\beta)$. If we define
\begin{IEEEeqnarray}{rLl}
\hat{S}_1(\beta) = \sum_{p,q} s^p_q(\beta) \left\lbrace \hat{p}^{\dagger} \hat{q}\right\rbrace \\ 
\hat{S}_2(\beta) = \frac{1}{4}\sum_{p,q,r,s} s^{pq}_{rs}(\beta) \left\lbrace \hat{p}^{\dagger} \hat{q}^{\dagger} \hat{s} \hat{r}\right\rbrace 
\end{IEEEeqnarray}
One obtains explicitly
\begin{IEEEeqnarray}{rLl}
D^p_q = \left\langle \hat{p}^{\dagger} \hat{q} (1 + \hat{S}_1) \right\rangle = f \delta^p_q +f \bar{f} s^q_p
\end{IEEEeqnarray}
\begin{IEEEeqnarray}{rLl}
D^{pq}_{rs} = \left\langle \hat{p}^{\dagger} \hat{q}^{\dagger} \hat{s} \hat{r} (1 + \frac{1}{2}\left\lbrace \hat{S}_1^2 \right\rbrace  + \hat{S}_2) \right\rangle &= ... = D^{p}_{r}D^{q}_{s} - D^{p}_{s}D^{q}_{r} + f^2 \bar{f}^2 s_{pq}^{rs}
\end{IEEEeqnarray}
It follows that the two-body amplitudes $s^{pq}_{rs}(\beta)$ represent the two-body cumulant up to a scaling factor. One can think of the normal-ordered exponential ansatz as a reduced density cumulant ansatz, and this is the perspective of Mukerjee and coworkers \cite{sanyal1992thermal,mandal1998thermal,mandal2001non,mandal2003finite}. Moreover, the above identification suggests that the operator $\hat{S}$ should be Hermitean. This aspect is not clear however, at the moment of writing this paper. It is possible that it is better for the operator $\hat{S}$ to remain non-Hermitean, and to obtain density matrices that are formally not necessarily Hermitean. Since operators that represent properties \emph{are} Hermitean, only the Hermitean part of the reduced density matrix would survive when takin gthe trace, and one can Hermitize after the fact. Alternatively, one could explicitly Hermitize the residual equations, and if we define 
\begin{IEEEeqnarray}{rLl}
\dot{\Omega}_{\nu}^+ = \frac{1}{2} \left( \dot{\Omega}_{\nu} + \dot{\Omega}_{\nu}^{\dagger} \right) 
\end{IEEEeqnarray}
the Hermitized residual equations are given by
\begin{IEEEeqnarray}{rLl} \label{herm_grand_amp}
- \left\langle   \dot{\Omega}_{\nu}^+  \frac{d\dot{S}}{d\beta} \right\rangle &= \left\langle  \dot{\Omega}_{\nu}^+   \left[ ( \dot{H}- \mu \dot{N}) \left\lbrace exp(\dot{S}(\beta)) \right\rbrace \right]_{connected} \right\rangle \\
-  \frac{ds_0}{d\beta} &= \left\langle   \left[ ( \dot{H}- \mu \hat{N}) \left\lbrace exp(\dot{S}(\beta)) \right\rbrace \right]_{connected} \right\rangle 
\end{IEEEeqnarray}
Curiously, nothing in the formal theory imposes this Hermitized form and it appears Hermitization is a choice, on which we remain agnostic at the moment. The issue will not be resolved in this paper. In particular in the simple model applications to be discussed in this paper, which serve mostly as  illustrations and sanity checks of the theory, the hermization does not do anything.  Let us note, however, that if we would consider the detailed singles and doubles equations we would observe a strong formal analogy between the present theory and conventional single reference coupled cluster, and it would appear that the non-Hermitean version of the theory is perhaps the better choice. In the formal development we will keep using projection against $\dot{\Omega}_{\nu}^+$ to emphasize the issue, but the actual preferred projection of choice will have to await more detailed numerical studies. 
  
The zero temperature, or $\beta \rightarrow \infty$ limit is obtained by setting the $\beta$ derivative (or left hand side of the amplitudes equations \Eref{eq_grand_amp}) to 0, hence, the pair of equations
\begin{IEEEeqnarray}{rLl} \label{grand_gs_amp}
 \left\langle  \dot{\Omega}_{\nu}^+   \left[ ( \dot{H}- \mu \dot{N}) \left\lbrace exp(\dot{S}(\beta)) \right\rbrace \right]_{connected} \right\rangle &=0 \\
 \left\langle  \left[ \dot{N} \left\lbrace exp(\dot{S}(\beta)) \right\rbrace \right] \right\rangle &= 0
\end{IEEEeqnarray}
defines a ground state methodology, which has many similarities with the contracted Schr\"{o}dinger equation or many-body cumulant theory. In particular there is no preferred set of occupied Hartree-Fock orbitals in the theory. It can be shown that the zero temperature limit yields exactly the same equations as the connected cumulant formulation by Nooijen et al \cite{nooijen2003cumulant}. We will not discuss this aspect further in this pedagogically oriented paper. The N-representability problem is a major issue with cumulant theory, and it is not unlikely this will also plague the present thermal or ground state theory.

The general formulation making extensive use of the concept of normal ordering and solving a differential equation is really quite nimble and elegant in our opinion. The zero temperature formulation can be tested in a straightforward way against full CI or other accurate results that have been obtained for many systems. Here the main difficulty perhaps lies in solving the equations, which may require special care. The testing of the finite temperature methodology is less straightforward as this problem has not been widely studied for molecular systems. The most straightforward way is to perform sum over states calculations (of varying number of electrons) in small finite basis sets, or restricting the problem to a complete active space (CAS). Magnetic model systems may be particularly suitable for this purpose. The $\beta$ dependent formulation is daunting perhaps, because one has to integrate the differential equation from $\beta=0$, where all electronic state are populated equally, to large values of $\beta$ where only few states carry significant population. Applications of the theory to all-electron situations with large one-particle basis sets may be challenging.

The reader may be surprised by the apparent power of the formulation and the concept of normal-ordering, e.g. when comparing the left and right hand side of the equation
\begin{IEEEeqnarray}{rLl} 
\frac{1}{Z} Tr(\hat{O} e^{-\beta \hat{H}}) &= \braket{\hat{O} \left \lbrace e^{\dot{S}(\beta)} \right \rbrace }
\end{IEEEeqnarray}
On the left hand side we have a very complicated expression even when $\hat{H}$ is a known two-body operator. On the right hand side we essentially have the solution when $\dot{S}(\beta)$ is known. The essential difference is those innocuous braces denoting normal ordering. There is no trace to be taken over a gigantic Hilbert space. The above equations provide essentially a recipe to calculate directly the thermal reduced density matrices. Powerful indeed. 
In the next section we will first perform some sanity tests for the theory. Our focus will be on one-electron Hamiltonians that are easily solved using simple computer implementations and concerns the case of non-interacting fermionic particles in the traditional grand canonical formulation, for which we expect to obtain the one-electron Fermi-Dirac result.

\section{\label{sec:elec-one}Detailed equations for grand canonical formulation using non-iteracting one-body hamiltonian}
Let us derive detailed equations for the one-body problem that should yield the usual solution given by the one-particle Fermi-Dirac partition function. The various operators in normal order are given by
\begin{IEEEeqnarray}{rLl}
\hat{h} &= h_0 + \sum_{p,q} h_{pq} \left\lbrace \hat{p}^\dagger \hat{q} \right\rbrace;  \qquad h_0 = f \sum_p h_{pp} \\
\hat{N} &= n_0 + \sum_{p,q} {\delta}_{pq} \left\lbrace \hat{p}^\dagger \hat{q} \right\rbrace;  \qquad n_0 = f \sum_p \delta_{pp} = n_{el} \\
\hat{S}(\beta) &= s_0(\beta) + \sum_{pq} s_{pq}(\beta) \left\lbrace \hat{p}^\dagger \hat{q} \right\rbrace ;    
\end{IEEEeqnarray}
Applying wick's theorem we can evaluate properties from the general formulas as 
\begin{IEEEeqnarray}{rLl}
U(\beta) &= h_0 + f\bar{f} \sum_{p,q} h_{pq}s_{qp}(\beta) \\
N_{el}&=n_{el} + f\bar{f} \sum_{pq} \delta_{pq}s_{qp}=n_{el} + f\bar{f} \sum_p s_{pp}(\beta) 
\end{IEEEeqnarray}
Since $N_{el}$ should evaluate to $n_{el}$ we obtain a constraint for the s-coefficients:
\begin{IEEEeqnarray}{rLl}
\sum_p s_{pp}=0
\end{IEEEeqnarray}
 The equation for $U(\beta)$ indicates also the proper expression for the one-particle reduced density matrix of the system.
\begin{IEEEeqnarray}{rLl}
D_{pq}(\beta) = f \delta_{pq} + f \bar{f} s_{pq}(\beta)
\end{IEEEeqnarray}
This quantity can direcly be compared with the exact result: If one diagonalizes the one-particle hamiltonian one obtains eigenvalues $\epsilon_p$, and the density matrix takes the diagonal Fermi-Dirac form  $n_p = \left[ 1 + exp(\beta (\epsilon_p - \mu)) \right] ^{-1}$. If one transforms this back to the original basis one obtains the density matrix which should agree with the expression for $\textbf{D}(\beta)$.

The amplitude equations are obtained from
\begin{IEEEeqnarray}{rLl}
- \left\langle \left\lbrace \hat{q}^\dagger \hat{p} \right\rbrace \frac{d\dot{S}}{d\beta} \right\rangle &= \left\langle \left\lbrace \hat{q}^\dagger \hat{p} \right\rbrace \left[ ( \dot{h}- \mu \dot{N}) \left\lbrace exp(\dot{S}(\beta)) \right\rbrace \right]_{connected} \right\rangle \\
-\frac{ds_0}{d \beta} &= \left\langle  \left[ \dot{h} \left\lbrace exp(\dot{S}(\beta)) \right\rbrace \right]_{connected}  \right\rangle - \mu n_{el}
\end{IEEEeqnarray}
 Substituting the detailed second quantized expressions and evaluating using Wick's theorem one finds
\begin{IEEEeqnarray}{rLl}
-f \bar{f} \frac{ds_{pq}}{d \beta} &= f \bar{f} \left( h_{pq} + \bar{f} \sum_r h_{pr} s_{rq} - f \sum_r s_{pr} h_{rq} - f \bar{f} \sum_{rs} s_{pr} h_{rs} s_{sq} \right) \nonumber \\
& \qquad -  \mu f \bar{f} \left( \delta_{pq} + \bar{f} s_{pq} - f s_{pq}  - f \bar{f} \sum_r  s_{pr} s_{rq} \right) \\
-\frac{ds_0}{d\beta} &= h_0 + f\bar{f} \sum_{p,q} h_{pq}s_{qp}(\beta) - \mu n_{el}
\end{IEEEeqnarray}
One can cancel a common factor $f \bar{f}$ from the amplitude equations and abbreviate the equations as
\begin{IEEEeqnarray}{rLl}
 - \frac{ds_{pq}}{d \beta} &= R_{pq}(h,s) - \mu R_{pq}(\delta,s) 
\end{IEEEeqnarray}
The parameter $\mu(\beta)$ is determined from the condition that the trace of $\textbf{s}$ should vanish, hence
\begin{IEEEeqnarray}{rLl}
\mu(\beta) &= \frac{\sum_p R_{pp}(h, s)}{\sum_p R_{pp}(\delta, s)} 
\end{IEEEeqnarray}
The differential equation can be solved in a simple numerical fashion using a leap frog scheme with a fixed step size $\Delta \beta$. 
\begin{IEEEeqnarray}{rLl}
S_{pq}(\beta_{i+1}) &= S_{pq}(\beta_{i-1}) + 2 \Delta \beta \frac{ds_{pq}}{d \beta}(\beta_i) \nonumber \\
&=  S_{pq}(\beta_{i-1}) - 2 \Delta \beta \left( R_{pq}(h,s(\beta_i)) - \mu(\beta_i) R_{pq}(\delta,s(\beta_i)) \right) \\
s_0((\beta_{i+1}) &=  s_0((\beta_{i-1}) + 2 \Delta \beta  \frac{ds_0}{d \beta}(\beta_i)
\end{IEEEeqnarray}

These equations have been implemented in a simple Fortran code, and it has been verified that the results reproduce the analytical results corresponding to Fermi-Dirac results obtained using a sum over states formulation and exact diagonalization of the finite basis hamiltonian. We used simple e.g. 20x20 hamiltonian matrices. It has also been verified that the natural occupation numbers do not become less than 0, nor do they exceed 1, and this is an interesting aspect, that deserves some further discussion.

The equations take on a simplified form in the basis in which $h$ is diagonal. Let us denote the orbital values as $\epsilon_p$. The operator $\hat{S}$ is diagonal, and we will simply refer to the amplitudes as $s_p$. The amplitude equations and natural occupuation numbers then read
\begin{IEEEeqnarray}{rLl}
- \frac{ds_{p}}{d \beta} &= (\epsilon_p-\mu) \left( 1 + \bar{f} s_p - f s_p - f \bar{f}  s^2_{p} \right)  \nonumber \\
&=(\epsilon_p-\mu) R(s_p) \\
R(s)&=1 + \bar{f} s - f s - f \bar{f}  s^2 \\
\frac{dR}{ds}&=\bar{f}-f -2 f \bar{f}  s \\
n_p &= f + f \bar{f} s_p 
\end{IEEEeqnarray}
The critical values for $s_p$ are
\begin{IEEEeqnarray}{rLl}
n_p=0 ;   s_p = -\bar{f}^{-1} \\
n_p=1 ; s_p = f^{-1}
\end{IEEEeqnarray}
We can see that at these extremal values we have
\begin{IEEEeqnarray}{rLl}
s=-\bar{f}^{-1}: R(s) = 0; \frac{dR}{ds}=1 \\
s = f^{-1}: R(s) = 0; \frac{dR}{ds}=-1
\end{IEEEeqnarray}
The meaning of these equations is that the rate of change $\frac{ds_p}{d \beta}$ goes to zero as the occupation numbers reach their extremal values. For these extreme values the value of $(\epsilon_p-\mu)$ will be far from zero, and so the sign of $\frac{ds_p}{d \beta}$ is uniform. As $\beta$ increases the values will approach the extremal values closer. As long as the integration step is not too large the occupation numbers will fall in the interval $\left[0,1 \right]$. 
The solution of the equations in the diagonal basis mimics the solution in the non-diagonal basis. One can simply transform all quantities back from the diagonal basis to the original basis, and this simply presents an overall unitary transformation of the equations. It follows that the density matrix is Hermitean for all values of $\beta$, without any need for explicit Hermitization of the equations. The issue of the need for Hermitization cannot be assessed from the above example.

At infinite $\beta$ (or zero temperature) all values of $s_p$ have reached their extremal values, and $\frac{ds_p}{d \beta}=0 \quad \forall p$. The ground state amplitudes and associated chemical potential can be determined directly from the equations  
\begin{IEEEeqnarray}{rLl}  \left\langle \left\lbrace \hat{q}^\dagger \hat{p} \right\rbrace \left[ ( \dot{h}- \mu \dot{N}) \left\lbrace exp(\dot{S}) \right\rbrace \right]_{connected} \right\rangle &=0 \\
\sum_p S_{pp}&=0
\end{IEEEeqnarray}
These conditions, in the diagonal basis, imply $R(s_p)=0 \quad \forall p$, or all occupation numbers take on their extremal value. Any such determinantal state satisfies the zero temperature equation. The ground state corresponds to the lowest energy solution. The thermal differential equations automatically converge to the ground state. The excited state solutions for the zero-temperature equations are additional solutions associated with the non-linearity of the equations.

In principle we could present some numerical results here but we learn nothing more than that the results agree exactly with one-electron Fermi-Dirac theory. This is a very satisfying, boring, result that proves the essential validity of the theory.

\section{\label{sec:vib-general} General Theory for pure bosonic systems}
We assume bosonic creation and annihilation operators that satisfy the usual commutation relations 
\begin{IEEEeqnarray}{rLl}
\hat{i} \hat{j}^{\dagger} - \hat{j}^{\dagger} \hat{i} &= \delta_{ij}
\end{IEEEeqnarray}
We define normal ordering and contractions 
\begin{IEEEeqnarray}{rLl}
\left\langle \hat{j}^{\dagger} \hat{i} \right\rangle =  f \delta_{ij}  ; \quad \left\langle \hat{i} \hat{j}^{\dagger}  \right\rangle &= \bar{f} \delta_{ij} \nonumber \\
\hat{j}^{\dagger} \hat{i} &= \left\lbrace \hat{j}^{\dagger} \hat{i} \right\rbrace + f \delta_{ij};  \quad \hat{i} \hat{j}^{\dagger}  = \left\lbrace \hat{i} \hat{j}^{\dagger}  \right\rbrace + \bar{f} \delta_{ij};  \nonumber \\
\bar{f} &= 1 + f  
\end{IEEEeqnarray}
Within a normal ordered product the operators commute, and the last relation can be derived from the commutation relations. The factor $f$ is as yet not defined. We assume here for simplicity that it is the same for all modes $i$, but the theory is easily generalized.

For definiteness we might be interested in vibrational problems and the second quantized operators would be associated with dimensionless normal mode coordinates $q_i$ in the usual way, based on an underlying quadratic potential around an equilibrium geometry. We will use a very similar second quantized approach to tackle both thermal properties and time-autocorrelation functions to gain access to spectroscopy.

The Hamiltonian is expressed in second quantization, and can be written in normal order, e.g. 
\begin{IEEEeqnarray}{rLl}
\hat{H} &= h_0 + \sum_i h^i  \left\lbrace \hat{i}^{\dagger} \right\rbrace + \sum_i h_i \left\lbrace \hat{i} \right\rbrace  +  \sum_{ij} h^i_j \left\lbrace \hat{i}^{\dagger} \hat{j} \right\rbrace +  \sum_{ij} h^{ij} \left\lbrace \hat{i}^{\dagger} \hat{j}^{\dagger} \right\rbrace +  \sum_{ij} h_{ij} \left\lbrace \hat{i} \hat{j} \right\rbrace + \cdots \nonumber \\
&= h_0 + \sum_{\lambda} h_{\lambda} \left\lbrace \hat{\Omega}_{\lambda} \right\rbrace 
\end{IEEEeqnarray}
The procedure we use to calculate thermal properties is very similar to the procedure discussed in the previous section. The main difference is that we do not include a constraint on the number of particles, and the use of commutation rather than anticommutation relations, of course. Given the form of the Hamiltonian we define an operator $\hat{S}(\tau)$ in a similar way
\begin{IEEEeqnarray}{rLl}
\hat{S} &= s_0 + \sum_{\lambda} s_{\lambda}(\tau) \left\lbrace \hat{\Omega}_{\lambda} \right\rbrace \equiv s_0 + \dot{S} 
\end{IEEEeqnarray}
The parameters that define the operator $\hat{S}$ are defined from the differential equation 
\begin{IEEEeqnarray}{rLl}
\left\langle \hat{\Omega}_{\mu}^{\dagger} \frac{d\hat{S}}{d \tau} \right\rangle &=  \frac{1}{k_B \tau^2} \left\langle \hat{\Omega}_{\mu}^{\dagger} \big( \hat{H} \left\lbrace e^{\hat{S}} \right\rbrace \big)_{connected} \right\rangle 
\end{IEEEeqnarray}
To evaluate the thermal expectation value of any operator $\hat{O}$ we evaluate 
\begin{IEEEeqnarray}{rLl}
\left\langle \left\langle \hat{O} \right\rangle \right\rangle &= \frac{\left\langle  \big( \hat{O} \left\lbrace e^{\hat{S}} \right\rbrace \big)_{connected} \right\rangle }{\left\langle  \left\lbrace e^{\hat{S}} \right\rbrace  \right\rangle }  \nonumber \\
&=\left\langle \big( \hat{O} \left\lbrace e^{\dot{S}} \right\rbrace \big)_{connected} \right\rangle
\end{IEEEeqnarray}
The quantity $e^{s_0}$ represents the partition function. Please note the use of the (non-constant) operator component $\dot{S}$ in the last equation.
We formulate the theory in terms of an integration over temperature ($\tau$) from $0_+$ to $T$. This means we have to specify the initial value of the amplitudes $s_{\lambda}(\tau=0_+)$. In addition we have to define the contraction $f$. Here the symbol $0_+$ indicates a small finite temperature different from 0. The initial condition for the bosonic theory is a little cumbersome. We cannot (easily) start from the infinite T limit as the occupation numbers of bosons have no upper limit. Moreover we cannot start exactly at T=0, as the initial $S$-amplitudes solving the ground state would yield vanishing residuals and hence the propagation cannot get started, related to the essential flatness at 0 K: all derivatives of the function $e^{\frac{1}{k_B T}}$ vanish at $T=0$. The solution might be to solve for a few excited states that have some population at small T, (e.g. T=20 K). We can calculate the thermal reduced density matrices for this ensemble and extract the corresponding $S$-amplitudes. We will provide an example based on a general multidimensional displaced harmonic oscillator in $\Sref{sec:thermal-ho}$.

If one integrates the equations using temperature as a variable some care will be needed to obtain a stable numerical procedure. The equations of interest are coupled ordinary differential equations that can be expressed as
\begin{IEEEeqnarray}{rLl}
\frac{dS_{\lambda}}{d \tau} = \frac{1}{k_B \tau^2} R_{\lambda}(\textbf{S}(\tau))
\end{IEEEeqnarray} 
If we would use a leap frog type algorithm where we integrate beween $\tau_{i-1}$ and $\tau_{i+1}$, we can approximate
\begin{IEEEeqnarray}{rLl}
S_{\lambda}(\tau_{i+1}) - S_{\lambda}(\tau_{i-1}) &= \int_{\tau_{i-1}}^{\tau_{i+1}} d \tau \frac{1}{k_B \tau^2} R_{\lambda}(\textbf{S}(\tau_i)) \nonumber \\
S_{\lambda}(\tau_{i+1}) &= S_{\lambda}(\tau_{i-1}) - (\frac{1}{k_B \tau_{i+1}} - \frac{1}{k_B \tau_{i-1}}) R_{\lambda}(\textbf{S}(\tau_i))
\end{IEEEeqnarray}
Here we assume the factor $\frac{1}{\tau^2}$ varies more quickly than $R_{\lambda}(\textbf{S}(\tau))$. More sophisticated integration schemes may have to be explored. These equations clearly have troubles near $\tau = 0$, indicating once again that the propagation should be started from $\tau_0=0_+$ and using a suitable set of initial $S$-amplitudes.

The theory can also be developed in real time. We would obtain the differential equation
\begin{IEEEeqnarray}{rLl}
i \left\langle \hat{\Omega}_{\mu}^{\dagger} \frac{d\hat{S}}{d\tau} \right\rangle &=   \left\langle \hat{\Omega}_{\mu}^{\dagger} \big( \hat{H} \left\lbrace e^{\hat{S}(\tau)} \right\rbrace \big)_{connected} \right\rangle 
\end{IEEEeqnarray}
Taking a Fourier transform of $s_0(\tau)$ would yield information on eigenvalues. Again one needs to provide a suitable set of initial amplitudes at $\tau = t_0$.

Given the definition of contractions, normal order and Wick's theorem the equations can also be formulated as operator equations, rather than through a projection. Hence we can write
\begin{IEEEeqnarray}{rLl}
 \frac{d\hat{S}}{d \tau}  &=  \frac{1}{k_B \tau^2}   \big( \hat{H} \left\lbrace e^{\hat{S}(\tau)} \right\rbrace \big)_{connected} 
\end{IEEEeqnarray} 
This equation is to be viewed as an equation for the amplitudes, where the left-hand and right-hand amplitudes are equated component by component. Let us note that the notion of connected in this context implies as usual that the expression is written in  normal ordering. The same type of alternative viewpoints have been discussed in the context of Coupled Cluster theory for electronic structure theory, and for example similarity transformed equation of motion coupled cluster is preferably discussed in this fashion (see e.g.  \cite{nooijen1996general, nooijen1999similarity} for pedagogical discussions). However, for single reference theory the alternate approaches of derivation lead to equivalent equations. The same is true here, as the projection by $\hat{\Omega}_{\mu}$ simply picks out a unique component and multiplies left and right hand of equation by the same combination of factors $f, \bar{f}$.

\section {\label{sec:vib-1d} Bosonic harmonic oscillator examples in 1d}
At this point the theory appears to be very general but also quite abstract. What would be appropriate definitions of $f$? Can one actually integrate the equations? In the thermal case could one instead use $\beta$ as the variable and start from the high $T$ or $\beta =0$ limit (or small $\beta$)? These are interesting questions and we will address some of them using exceedingly simple but instructive examples. To illustrate the thermal case we will look at simple harmonic oscillators, starting with a one-dimensional problem before discussing the general multidimensional problem. For time-dependent problems we will formulate a procedure to calculate time-autocorrelation functions for Franck-Condon spectra based on Harmonic oscillators for ground and excited states.

In this sctione will discuss the simple example of a one-dimensional (undisplaced) harmonic oscillator.  The equations are derived in detail, and the integration can be done analytically.

\begin{IEEEeqnarray}{rLl}
\hat{H} &= \omega \big( \hat{i}^{\dagger} \hat{i} + \frac{1}{2} \big) = h_0 +  \omega \left\lbrace \hat{i}^{\dagger} \hat{i} \right\rbrace ;   h_0 = (\frac{1}{2}+f) \omega
\end{IEEEeqnarray}
Defining the operator $\hat{S}$ in a similar way we obtain equations
\begin{IEEEeqnarray}{rLl}
\left\langle \left\lbrace \hat{i} \hat{i}^{\dagger} \right\rbrace \frac{d \hat{S}}{d \tau} \right\rangle &=  \frac{1}{k_B \tau^2} \left\langle \left\lbrace \hat{i} \hat{i}^{\dagger} \right\rbrace \big( \hat{H} \left\lbrace e^{\hat{S}} \right\rbrace \big)_{connected} \right\rangle \nonumber \\
\frac{ds_0}{d \tau} &=  \frac{1}{k_B \tau^2} \left\langle \big( \hat{H} \left\lbrace e^{\hat{S}} \right\rbrace \big)_{connected} \right\rangle
\end{IEEEeqnarray}
which can be evaluated as
\begin{IEEEeqnarray}{rLl}
\label{eqn-1d-thermal-s}
f \bar{f} \frac{ds}{d \tau} &=  \frac{1}{k_B \tau^2} \omega  \big( f \bar{f} + f \bar{f} (f + \bar{f}) s + f \bar{f} f \bar{f} s^2) \nonumber \\
\frac{ds_0}{d \tau} &=  \frac{1}{k_B \tau^2} (h_0 + \omega f \bar{f}  s)
\end{IEEEeqnarray}
If we use that the operator $\hat{S}$ is always contracted in the same manner we can substitute $t = f \bar{f} s$ (as we did in the section on the fermionic case) and obtain
\begin{IEEEeqnarray}{rLl}
 \frac{dt}{d \tau} &=  \frac{1}{k_B \tau^2} \omega  \big( f \bar{f} +  (f + \bar{f}) t +  t^2) = \frac{1}{k_B \tau^2} \omega  (t+f)(t+\bar{f}) \nonumber \\
 \frac{ds_0}{d \tau} &=  \frac{1}{k_B \tau^2} \omega (\frac{1}{2} + f + t)
\end{IEEEeqnarray}
The differential equation for $s$ can be solved analytically
\begin{IEEEeqnarray}{rLl}
 \frac{dt}{(t+f)(t+\bar{f})} &=  \frac{1}{k_B \tau^2} \omega  d \tau \nonumber \\
 \frac{dt}{(t+f)} - \frac{dt}{(t+\bar{f})} &=   \frac{1}{k_B \tau^2} \omega  d \tau 
\end{IEEEeqnarray}
This equation can be integrated, replacing $\tau$ by T, and introducing an integration constant $C$, which we set to zero quickly, using hindsight
\begin{IEEEeqnarray}{rLl}
\frac{(t+f)}{(t+\bar{f})}= e^{- \frac{\omega }{k_B T}}  + C \nonumber \\
t+f = e^{- \frac{\omega }{k_B T}}  (t+1+f) \nonumber \\
(t+f)( 1-e^{- \frac{\omega }{k_B T}} ) &= e^{- \frac{\omega }{k_B T}} \nonumber \\
t+f &= \frac{e^{- \frac{\omega }{k_B T}}}{1-e^{- \frac{\omega }{k_B T}}} \nonumber \\
t+f &= \frac{1}{e^{ \frac{\omega }{k_B T}}-1} 
\end{IEEEeqnarray}
The partition function can be obtained by integrating the corresponding equation for $\frac{ds_0}{d \tau}$.
The expectation value of any operator $\hat{O} = o \hat{i}^{\dagger} \hat{i} $ is evaluated as 
\begin{IEEEeqnarray}{rLl}
\left\langle \left\langle \hat{O} \right\rangle \right\rangle 
&=\left\langle \big( \hat{O} \left\lbrace e^{\dot{S}} \right\rbrace \big)_{connected} \right\rangle
&= o (f + f \bar{f} s) = o (t+f) = o \frac{1}{e^{ \frac{\omega }{k_B \tau}}-1}
\end{IEEEeqnarray}
This is the expected result, and $t+f$ represents the thermal density matrix. The above equations indicate that the contraction $f$ is essentially arbitrary. The initial value for $t$ can be taken to be $-f$, while the initial value for $s_0 = - \frac{1}{2} \omega / (k_B T)$, related to the ground state energy. While the solutions for the probablem can be obtained analytically they also indicate that the initial conditions for the bosonic problem are not trivial, and at low $T$ the amplitudes diverge or grow very slowly due to the behaviour of the Boltzmann-factor.  These issues will be addressed in subsequent sections. The above simple calculation indicates that we can expect the essential theory to work as advertised. As discussed more extensively below one can anticipate exact results for any Hamiltonian that is harmonic (or even more general: at most quadratic in annihilation/creation operators). Presumably the Hamiltonian does not even have to be Hermitean. As long as the parameters defining the Hamiltonian are real, the $t$-amplitudes and reduced density matrices are real too. 

Let us re-emphasize that the essential result is independent of the factor $f$. The value of $t$ depends on $f$, but the physically relevant quantity is $n=f+t$, the reduced density matrix (or population in the 1-d case) is independent of $f$. However, if one assumes (perhaps most naturally) that $f=0$, using the traditional ground state of the harmonic oscillator to define normal ordering, some unexpected issues do arise. If we look at the equation for amplitudes $s$, \Eref{eqn-1d-thermal-s} and divide first by $f \bar{f}$ on both sides, one obtains, setting $f=0$ afterwards,
\begin{IEEEeqnarray}{rLl}
 \frac{ds}{d \tau} &=  \frac{1}{k_B \tau^2} \omega  \big( 1+ (f + \bar{f}) s +  f \bar{f} s^2) \big ) \nonumber \\
 &= \frac{1}{k_B \tau^2} \omega (1+\bar{f} s) = \frac{1}{k_B \tau^2} \omega (1+s)
\end{IEEEeqnarray}
The right hand side is now linear in $s$ and has as a solution the Boltzman distribution $1+s(\tau) = e^{- \frac{\omega}{k_B \tau}}$. Incidentally, the exact same equation is obtained if one uses the many-body formulation. Clearly this result, setting $f=0$ is not correct . Rather one would have to solve the equation for non-zero $f$ and then take $lim{f \rightarrow 0}$ in the end result. This result would be correct, and the result are identical as before if one substitutes $t=f \bar{f} s$. However, if one examines the equations for $t$, one can take the limit before integrating, yielding
\begin{IEEEeqnarray}{rLl}
 \frac{dt}{d \tau} &=  \frac{1}{k_B \tau^2} \omega  \big(   \bar{f} t +   t^2) \big ) \nonumber \\
 &= \frac{1}{k_B \tau^2} \omega (t + t^2)
\end{IEEEeqnarray}
This differential equation  has the solution $t(\tau) = \frac{1}{e^{ \frac{\omega}{k_B \tau}} - 1}$, which \emph{is} correct. The situation is delicate because $f \bar{f} s = t$ is finite, even in the limit $f \rightarrow 0$. In summary, to derive the equations we need to use $\hat{S}$. In the case that we use $f \neq 0$ one can use either many-body or projected equations. However, when we use $f=0$, \emph{only} the projected equations will work \emph{and only} if we first make the substitution $f \bar{f} s \rightarrow t$. It might seem most pertinent to avoid the case $f=0$. However, in the case of solving the Franck-Condon problem the use of $f=0$ is most convenient. The procedure of deriving equations using $s$ and then redefining variables to introduce $t$ yields the simplest equations in practice.  Let us note that the case $f=0$ is very natural as it corresponds to a legitimate normal ordering in which annihilation operators are always to the left of creation operators. It is the normal ordering used for example in work by Facheaux and Hirata, \cite{faucheaux2015higher,faucheaux2017diagrammatic,faucheaux2018similarity}, and is commonly used in conjunction with second quantization for vibrational problems. The above analysis shows it may actually (rather easily) lead to erroneous equations, at least in the thermal case.

\section{\label{sec:thermal-ho} Explicit amplitude equations for thermal equations for Bosons in a singles and doubles approximation}
Let us next discuss the general problem for the multidimensional Harmonic oscillator addressing a numerical way to find suitable initial conditions for the amplitudes. The results are compared numerically with sum over states results and we again achieving the satisfying, but boring result that everything works as advertised. 

The harmonic oscillator Hamiltonian expressed in normal order (assuming some $f$) is given by
\begin{equation}
\hat{H}=h_0+\sum_ih_{i}\hat{a}_i^\dagger+\sum_ih^{i}\hat{a}_i+\sum_{ij}h^i_j \{ \hat{a}_i^\dagger\hat{a}_j \} + \frac{1}{2}\sum_{ij}h^{ij}\hat{a}_i^\dagger\hat{a}_j^\dagger + \frac{1}{2}\sum_{ij}h_{ij}\hat{a}_i\hat{a}_j 
\end{equation}
where $h_0=E_0 + f \bar{f} \sum_i  \omega_i$, where $E_0$ is the ground state energy (including the zero point energy) of the harmonic oscillator. Likewise the operator $\hat{S}$ is parameterized as
\begin{equation}
\label{CC_operator}
\hat{S}=s_0+\sum_is_{i}\hat{a}_i^\dagger+\sum_is^{i}\hat{a}_i+\sum_{ij}s^i_j \{ \hat{a}_i^\dagger\hat{a}_j \} + \frac{1}{2}\sum_{ij}s^{ij}\hat{a}_i^\dagger\hat{a}_j^\dagger + \frac{1}{2}\sum_{ij}s_{ij}\hat{a}_i\hat{a}_j 
\end{equation}
The equations are derived by evaluating the following equations through an application of Wick's theorem and keeping only fully contracted terms
\begin{equation}
-\langle \hat{\Omega}_{\lambda}^{\dagger} \frac{d\hat{S}}{d\beta} \rangle = \langle \hat{\Omega}_\lambda^\dagger \big( \hat{H} \left \lbrace e^{\hat{S}} \right \rbrace \big)_{connected}\rangle
\end{equation}
Here the operator manifold is given by 
\begin{equation}
\hat{\Omega}_\lambda \equiv \{\hat{1}, \hat{a_i}, \hat{a}_i^\dagger , \hat{a}_i\hat{a}_j, \hat{a}_i^\dagger \hat{a}_i^\dagger, \left\lbrace \hat{a}_i^\dagger \hat{a}_j\right\rbrace\}
\end{equation}

Since the expressions are always fully contracted, each $s$ amplitude always carries the corresponding factors of $f, \bar{f}$. In addition the operators on the Hamiltonian that are contracted with the external projection operators carry corresponding factors $f$ and $\bar{f}$. We replace the $s$-amplitudes by $t$-amplitudes using the completely systematic substitutions $f s^i \rightarrow t_i, \bar{f} s^i \rightarrow t^i, f^2 s^{ij} \rightarrow t^{ij}, \bar{f}^2 s_{ij} \rightarrow t_{ij}, f \bar{f} s^i_j \rightarrow t^i_j$. An external creation operator (upper index) on $h$ retains a factor $\bar{f}$, while lower external indices on $h$ carry a factor $f$ from the original evaluation of Wick's theorem. These rules (that are easy to prove) facilitate the evaluation of terms. The detailed results are given below. Note that we use Einstein summation to express the sums over the labels of the tensors. We obtain the following.

Zero order equation:
\begin{equation}
\begin{split}
-\frac{ds_0}{d\beta} &= h_0+h_k^lt^k_l+h^l_kt^kt_l+h_kt^k+h^kt_k \\
&+\frac{1}{2}h_{kl}t^{kl}+\frac{1}{2}h_{kl}t^kt^l+\frac{1}{2}h^{kl}t_{kl}+\frac{1}{2}h^{kl}t_kt_l
\end{split}
\end{equation}
Single equations:
\begin{equation}
\begin{split}
-\frac{dt^i}{d\beta} &= \bar{f}h^i + h^kt_k^i + \bar{f}h^i_kt^k + h^k_l t^i_kt^l + h^k_lt_kt^{li} + h^{kl}t_kt_{l}^i \\
&+ h_{kl}t^kt^{li} + h_kt^{ki} + h^{ik}t_k
\end{split}
\end{equation}
\begin{equation}
\begin{split}
-\frac{dt_i}{d\beta} &= fh_i + h_kt^k_i + fh^k_it_k + h^l_kt^k_it_l + h^l_kt^kt_{li} + h^{kl}t_kt_{li} \\
&+ h_{kl}t^lt^k_i + h^kt_{ki} + fh_{ki}t^k
\end{split}
\end{equation}
Double equations
\begin{equation}
\begin{split}
-\frac{dt^i_j}{d\beta} &= f\bar{f}h^i_j + fh^k_jt^i_k + \bar{f} h^i_kt^k_j + h^k_lt^i_kt^l_j + h^k_lt_{kj}t^{li} \\
&+\bar{f}h^{ki}t_{kj} + h^{kl}t_{lj}t^i_k + h_{kj}t^{ik} + h_{kl}t^{ki}t^l_j
\end{split}
\end{equation}

\begin{equation}
\begin{split}
-\frac{dt^{ij}}{d\beta} &= h^{kl}t^i_kt^j_l + h_{kl}t^{ki}t^{lj} + h^k_lt^{li}t^j_k + h^k_lt^{lj}t^i_k \\
&+ \bar{f}h^j_kt^{ki} + \bar{f}h^i_kt^{kj} + \bar{f}h^{kj}t^i_k + \bar{f}h^{ki}t^j_k + \bar{f}\bar{f}h^{ij}  
\end{split}
\end{equation}

\begin{equation}
\begin{split}
-\frac{dt_{ij}}{d\beta} &= h^k_lt^l_it_{kj} + h^k_lt^l_jt_{ki} + h^{kl}t_{li}t_{kj} + h_{kl}t^k_it^l_j \\
&+ fh^k_it_{kj} + fh^k_jt_{ki} + fh_{ki}t^k_j + fh_{kj}t^k_i + ffh_{ij}
\end{split}
\end{equation}

The thermal quantities could be determined through the $\beta$ or  imaginary time integration over the amplitudes. The primary remaining issue is the determination of initial conditions. If we define the thermal density matrices at a particular value of $\beta$, as well as the partition function $Z=\sum_nu e^{-\beta E_{nu}}$ using a small sum over states expression, valid at low temperature (large ($\beta$)
\begin{equation}
d_{\lambda} = \frac{1}{Z}\sum_{\nu} \bra{\Psi_{\nu}} \hat{\Omega}_{\lambda} ^{\dagger} \ket{\Psi_{\nu}} e^{- \beta E_{\nu}}
\end{equation}
One can extract the initial conditions for the propagation (valid for any $f$).
\begin{equation}
\begin{split}
&s_0 = ln(Z) \\
&t_i = d^i \\
&t_i = d_i   \\
& t^{i}_{j} = d^i_j - t^{i}t_{j} - f \\
& t^{ij} = d^{ij} -t^i t^j \\
& t_{ij} = d_{ij} -t_i t_j \\
\end{split}
\end{equation}

Upon integration, thermal properties can be obtained. Most notably the partition function $Z$ and Helmholtz Free energy $A = - k_B T ln Z$
\begin{equation}
Z(\beta) = Tr(D) = e^{s_0(\beta)}
\end{equation}
as well as the thermal internal energy U and hence entropy ($TS = U-A$)
\begin{equation}
U = Tr(\hat{H}\hat{D}) = (\hat{H}e^{\hat{S}})_{f.c.} = -\frac{ds_0}{d\beta}
\end{equation}

The heat capacity can be obtained using differentation of $U$, if desired. Also thermal density matrices and cumulants are easily obtained using the (inverse) of the formulas above, relating density matrices and $t$-amplitudes.

Let us emphasize one more aspect of interest. For quadratic hamiltonians (containing up to two annihilation and/or creation operators) these propagation equations are exact (up to numerical precision of the integration, and depending on the validity of initial conditions). The reason is that the ansatz is in principle exact, but using up to two-body amplitudes the three-body results are zero as $\big( \hat{H} \left \lbrace e^{\hat{S}} \right \rbrace \big)_{connected}$ cannot lead to three-body terms under these conditions. 

Of course by itself these results are not very interesting. The approach is expected to be a good approximation also for more complicated hamiltonians. The main drawback is that the initialization of the thermal density matrices would currently require an external source for a few low lying states. It is to be noted that low lying states are generally more easily obtained than higher lying states, e.g. harmonic oscillator may be sufficiently accurate, or a perturbative approach can be adopted to include anharmonic effects. It is interesting that using the thermal NOE approach one recovers the complete, basis set free, partition function even when the starting point only includes a few states that have significant thermal population at low T. This latter aspect is not obvious form the theoretical considerations alone, and so it will be good to provide evidence through a numerical example. 

\begin{figure}[hbtp]
\centering
\subfloat[Partition function]{
  \includegraphics[scale = 0.5]{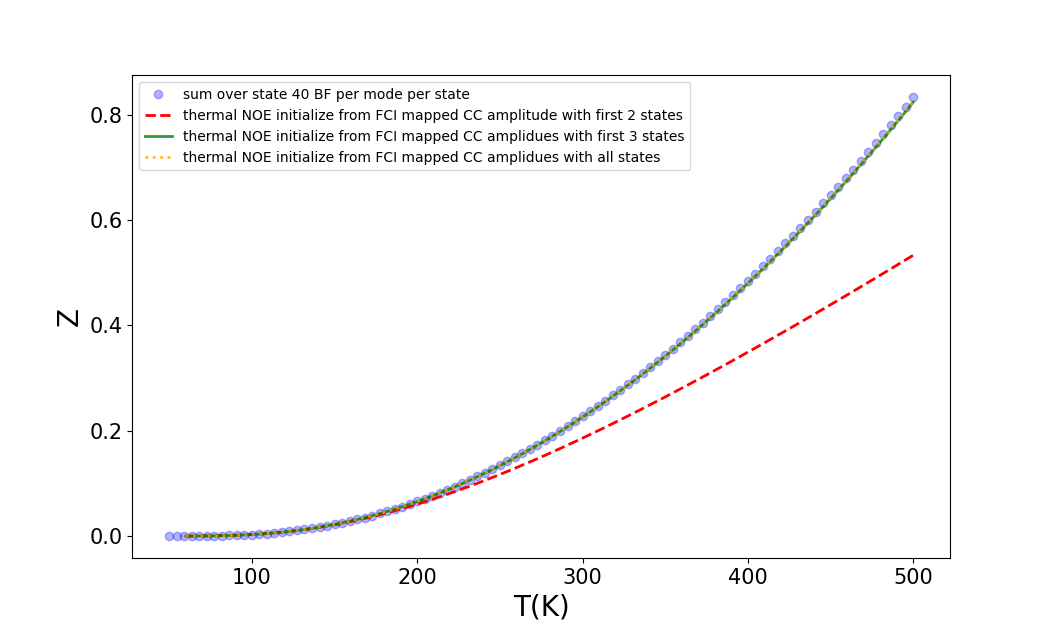}
}
\hspace{0mm}
\subfloat[Internal energy]{
  \includegraphics [scale = 0.5]{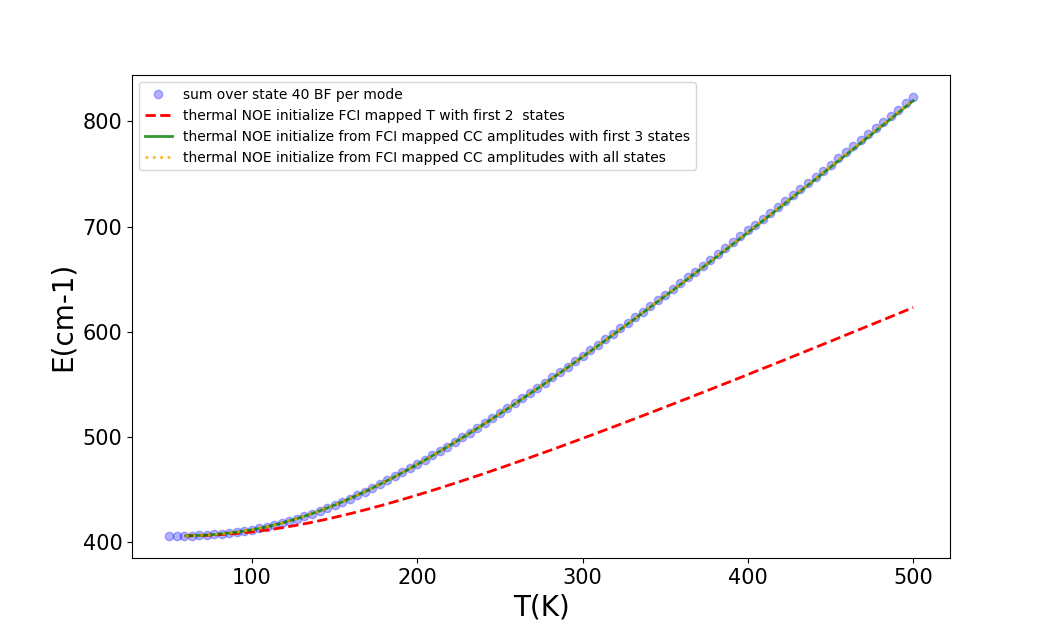}
}
\caption{Comparison of thermal NOE and sum over states approach to thermal properties for 2d harmonic oscillator . The thermal NOE is initialized using the thermal density for either two states (red) or 3 states (green). The latter matches the sum over states results. a) Partition Function. b) Internal energy.}
 
\label{fig:thermal_ho}
\end{figure}

We implemented the schemed described  above and tested the scheme for a general harmonic oscillator hamiltonian with two vibrational modes (corresponding to vibrational frequencies of 300 and 360 $cm^{-1}$. Using a small number of states (2 or 3 for comparison), obtained from a finite basis full CI calculation we extracted the thermal densities at the initial temperature of 60K and the propagated the thermal NOE equations up to 500K. The results for the partition function $Z$ and the internal energy $U$ are compared to the complete sum over states finite basis set calculation. It is easily seen from Fig. \Fref{fig:thermal_ho}  that the result are indeed numerically very accurate if the initial set of states is large enough. Including the three lowest vibrational states suffices to obtain the full partition function, while including two states is clearly not enough. At higher T we anticipate the (basis set free) NOE result to remain accurate while the full CI result will (eventually) suffer from the finite basis set used. 

\section{\label{sec:FC-spectra} Similation of harmonic Franck-Condon spectra}

It is clear that solving the thermal differential equations is not the preferred way for solving the simple pedagogical problems discussed thus far (one-electron problems for Fermions, harmonic oscillator for Bosons). Let us turn next to a problem that might be of actual interest to the quantum chemistry community: the simulation of harmonic Franck-Condon spectra using time-autocorrelation functions using the present NOE approach.

The math of time-dependent formalism resembles the thermal density matrix formalism but with different goals. In this case, our goal is to solve time dependent Sch\"odinger equation by simply switching the imaginary time integration in the thermal case to real time integration. We will here demonstrate the theory using the simple example of the calculation of a vibronic absorption spectra in the vertical Hessian harmonic Franck-Condon approximation. We assume the initial ground state wave function is described by the ground state of a harmonic oscillator, while the electronic transition moments are constant. The second quantized operators are the dimensionless normal modes of the ground state and we use as a vacuum state for Wick's theorem the corresponding ground vibrational state $\ket{0}$, such that the only non-zero contractions are given by 
\begin{equation}
\overline{\hat{i} \hat{i}^{\dagger} } = \bra{0} \hat{i} \hat{i}^{\dagger} \ket{0} = \bar{f} = 1;  
\end{equation}
while $f=\overline{\hat{i}^{\dagger}  \hat{i}} =0$
To obtain the spectrum we calculate the autocorrelation function based on the excited state harmonic hamiltonian, shifted by the ground state energy (vertical excitation energy plus ground state zeropoint frequency). This excited  state Hamiltonian is represented as usual
\begin{equation}
\hat{H}=h_0+\sum_ih_{i}\hat{a}_i^\dagger+\sum_ih^{i}\hat{a}_i+\sum_{ij}h^i_j \{ \hat{a}_i^\dagger\hat{a}_j \} + \frac{1}{2}\sum_{ij}h^{ij}\hat{a}_i^\dagger\hat{a}_j^\dagger + \frac{1}{2}\sum_{ij}h_{ij}\hat{a}_i\hat{a}_j 
\end{equation}

The time dependent Sch\"odinger equation(TDSE) we wish to solve is given by

\begin{equation}
i\frac{d}{d\tau}\mid\Psi(\tau)\rangle = \hat{H}\mid\Psi(\tau)\rangle
\end{equation}

If we apply the normal-ordered exponential ansatz to parameterize the time dependent wavefunction, we have

\begin{IEEEeqnarray}{rLl}
\ket{\Psi(\tau)} &= \left \lbrace e^{\hat{S}} \right \rbrace \ket{\Psi(\tau=0)} \nonumber \\
&= \left \lbrace e^{\hat{S}} \right \rbrace \ket{0}
\end{IEEEeqnarray}

Due to the normal-ordered exponential and the fact that $\hat{i} \ket{0} = 0$, it follows that the only non-trivial contributions in $\hat{S}$ contain creation operators only and such operators all commute, such that we do not need the normal ordering but can parameterize the wave function as
\begin{IEEEeqnarray}{rLl}
\ket{\Psi(\tau)}  &=  e^{\hat{T(\tau)}}  \ket{0}  \nonumber \\
\hat{T} &= t_0(\tau) + \sum_i t^i(\tau) \hat{i}^{\dagger}  + \frac{1}{2} \sum_{ij} t^{ij} \hat{i}^{\dagger} \hat{j}^{\dagger}
\end{IEEEeqnarray}

We can obtain suitable working equations by subsituting the ansatz into the time-dependent schr{\"o}dinger equation, multiplying by $e^{-\hat{T}}$ and projecting again $\bra{0} \hat{\Omega}_{\lambda}$
\begin{IEEEeqnarray}{rLl}
i \bra{0} \hat{\Omega}_{\lambda} \frac{d\hat{T}}{d \tau} \ket{0} &= \bra{0} \hat{\Omega}_{\lambda} e^{-\hat{T}} \hat{H} e^{\hat{T}} \ket{0} \nonumber \\
&= \bra{0} \hat{\Omega}_{\lambda}  \big( \hat{H} e^{\hat{T}}\big)_{connected} \ket{0} 
\end{IEEEeqnarray}

It can be seen that in this case the normal ordered ansatz reduces to traditional Coupled Cluster equations for vibrational problems. Using the projection manifold $\hat{\Omega}_\lambda = \{1,\hat{i}_i, \hat{i}_i \hat{j} \}$, we obtain CC amplitude equations as follow:

\begin{equation}
\begin{split}
i\frac{ds_0}{d\tau} &= h_0+h_k^lt^k_l+h^l_kt^kt_l+h_kt^k+h^kt_k \\
&+\frac{1}{2}h_{kl}t^{kl}+\frac{1}{2}h_{kl}t^kt^l+\frac{1}{2}h^{kl}t_{kl}+\frac{1}{2}h^{kl}t_kt_l
\end{split}
\end{equation}
\begin{equation}
\begin{split}
i\frac{dt^i}{d\tau} &= \bar{f}h^i + h^kt_k^i + \bar{f}h^i_kt^k + h^k_l t^i_kt^l + h^k_lt_kt^{li} + h^{kl}t_kt_{l}^i \\
&+ h_{kl}t^kt^{li} + h_kt^{ki} + h^{ik}t_k
\end{split}
\end{equation}
\begin{equation}
\begin{split}
i\frac{dt^{ij}}{d\tau} &= h^{kl}t^i_kt^j_l + h_{kl}t^{ki}t^{lj} + h^k_lt^{li}t^j_k + h^k_lt^{lj}t^i_k \\
&+ \bar{f}h^j_kt^{ki} + \bar{f}h^i_kt^{kj} + \bar{f}h^{kj}t^i_k + \bar{f}h^{ki}t^j_k + \bar{f}\bar{f}h^{ij}  
\end{split}
\end{equation}

At time $\tau=0$ all cluster amplitudes are zero and it is a simple matter to propagate the equations in time. The time-autocorrelation function $ACF(\tau)$ is given by
\begin{equation}
ACF(\tau) \equiv \langle \Psi(0)\mid\Psi(\tau\rangle = e^{s_0}
\end{equation}
One can obtain zero temperature Harmonic Franck-Condon spectra by taking the Fourier transform of $ACF(\tau)$ and multiplying by the square of the electronic transition moment. 
In the literature there are a large number of methods available to calculate harmonic franck-Condon spectra, including hot bands and Hertzberg-Teller effects. The above time-dependent CC approach yields numerically exact results for the simplest type of spectra and is perhaps the simplest approach among them. It appears possible to generalize the approach to i) more complicated surfaces, ii) hot bands, iii) Hertzberg teller effects. In our group we are aiming to generalize to approach to multistate vibronic models including non-adiabatic coupling. As an example we simulated the photo-electron spectrum of the formaldehyde molecule, comparing the CC result to an MCTDH calculation in a large single particle basis set. The excited state hamiltonian was obtained many years ago through an IP-EOMCC calculation. Here our goal is to show such calculations can indeed be done routinely, and they take mere minutes (or seconds) of computation time. The number of parameters is identical to a full Gaussian wave packet calculation.

\begin{figure}[hbtp]
\centering

\includegraphics[width = \linewidth]{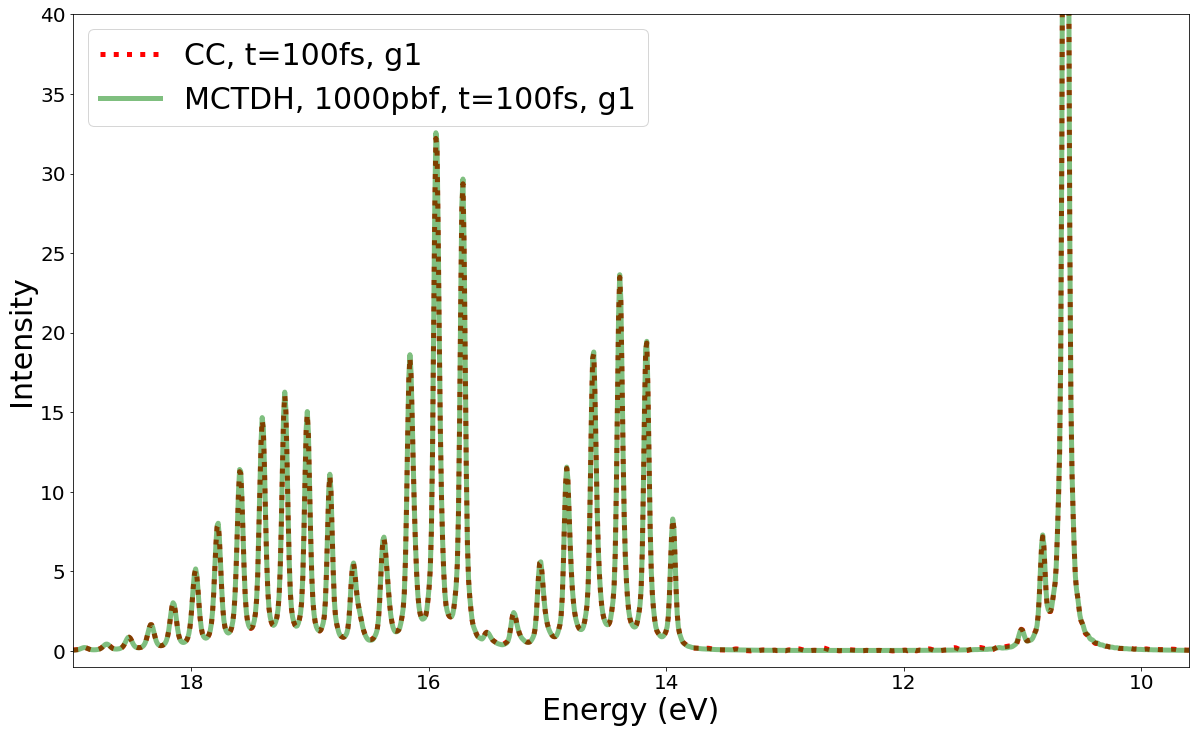}

\caption{Comparison of harmonic Franck-Condon photo-electron spectrum of formaldehyde:  Time-dependent NOE (this work) vs. converged MCTDH result}
 
\label{fig:FC-spectrum-h2co}
\end{figure}

\section{\label{sec:reflections} Further Reflections}
The above formulation of the calculation of thermal properties, time-dependent auto-correlation functions from solving coupled differential equations shows great unity between the approaches discussed for electronic structure theory and for pure vibrational problems. The basic difference between approaches discussed refers to the use of anti-commutation relations for Fermions, and commutation relations for Bosonic problems, and the associated relations $\bar{f} = 1 \pm f$ . Moreover, in electronic problems we have introduced the chemical potential $\mu$ to ensure the average electron count. Finally, in electronic structure problems we have found it convenient to start from the infinite temperature limit ($\beta=0$), while for Bosonic problems the more convenient starting point is $T=0_+$, and solving for the ground and low-lying excited states first, and extract initial values for the amplitudes from the associated thermal reduced density matrices. We have seen in the bosonic case that if one uses the (natural) value $f=0$, using the true vacuum, the equations should be phrased in terms of the 'renormalized' amplitudes \textbf{t}. Let me collect the relevant equations here, both in terms of $s$ and $t$ amplitudes for the simplest type of problem, to show the basic unity.

Bosons, including $s \rightarrow t$ substitution:
\begin{IEEEeqnarray}{rLl}
f \bar{f} \frac{ds}{d \tau} &= \frac{\omega}{k_B \tau^2} f \bar{f} \big(1 + (f + \bar{f}) s + f \bar{f} s^2 \big) \nonumber \\
 \frac{dt}{d \tau} &= \frac{\omega}{k_B \tau^2} \big( f \bar{f}  + (f + \bar{f}) t +  t^2 \big) \nonumber \\ 
 \lim{f \rightarrow 0} &: \frac{dt}{d \tau} = \frac{\omega}{k_B \tau^2} \big( t +  t^2 \big)
\end{IEEEeqnarray}
With bosonic solution (Bose-Einstein distribution)
\begin{IEEEeqnarray}{rLl}
n(\tau)&=(t+f)(\tau) = \frac{1}{e^{\frac{\omega}{k_B \tau}}-1  }
\end{IEEEeqnarray}
Fermions, including $s \rightarrow t$ substitution
\begin{IEEEeqnarray}{rLl}
f \bar{f} \frac{ds_i}{d \tau} &= \frac{\epsilon_i-\mu}{k_B \tau^2} f \bar{f} \big(1 + (\bar{f} - f) s_i - f \bar{f} s_i^2 \big) \nonumber \\
 \frac{dt_i}{d \tau} &= \frac{\epsilon_i-\mu}{k_B \tau^2} \big( f \bar{f}  + (\bar{f} -f) t_i - t_i^2 \big) \nonumber \\ 
 \lim{f \rightarrow 0} &: \frac{dt_i}{d \tau} = \frac{\epsilon_i-\mu}{k_B \tau^2} \big( t_i -  t_i^2 \big)
\end{IEEEeqnarray}
With Fermionic solution (Fermi-Dirac distribution)
\begin{IEEEeqnarray}{rLl}
n_i(\tau)&=(t_i+f)(\tau) = \frac{1}{e^{\frac{\epsilon_i-\mu}{k_B \tau}}+1  }
\end{IEEEeqnarray}
Boltzman (f=0 limit of  Bosonic  equation without $s \rightarrow t$ substitution):
\begin{IEEEeqnarray}{rLl}
f \bar{f} \frac{ds}{d \tau} &= \frac{\omega}{k_B \tau^2} f \bar{f} \big(1 + (f + \bar{f}) s + f \bar{f} s^2 \big) \nonumber \\
 \frac{ds}{d \tau} &= \frac{\omega}{k_B \tau^2}  \big(1 + (f + \bar{f}) s + f \bar{f} s^2 \big) \nonumber \\  
 \lim{f \rightarrow 0} &: \frac{ds}{d \tau} = \frac{\omega}{k_B \tau^2} \big( 1+s \big)
\end{IEEEeqnarray}
With  solution (Boltzman distribution)
\begin{IEEEeqnarray}{rLl}
n(\tau)&=(1+s)(\tau) = e^{\frac{-\omega}{k_B \tau}} 
\end{IEEEeqnarray}
The same Boltzmann distribution is obtained if one uses the same $f=0$, no $s \rightarrow t$ substitution in the Fermionic case. The Boltmann distribution (linear equation) is also obtained if one uses a Hamiltonian matrix (no second quantization).

All of these equations have the fundamental numerical problem that if they are initiated by first obtaining the ground state solution, and subsequently a first-order scheme is used to numerically integrate, the solution is dead as the derivatives at $T=0$ are all equal to zero, and the amplitudes from a first-order numerical integration scheme will remain zero.  As was shown before the diagonal one-body problems can easily be solved analytically, but numerical approaches, needed in practice, may suffer.

We think it is insightful to obtain the three fundamental statistics in Physics related to these simple linear or quadratic differential equations, where the difference in statics relates to the factor of the quadratic term. A factor of $+1$ yields Bose-Einstein Statistics, A factor of $-1$ yields Fermi-Dirac statistics, while a factor of $0$ yields Boltzmann statistics.  
Most surprisingly is that we are not aware of this discussion in the literature. It seems likely that it has been discussed before but we authors are not aware of it. The same notion is of course there if we write the general formula $n_i^{\alpha}(\tau)=(e^{\omega_i \tau} + \alpha)^{-1}$, 
where $\alpha = \pm 1, 0$ yields the various statistics. The novelty is the existence of the differential equations, which allows the generalization to more complicated Hamiltonians.

\subsection{\label{sec:comp-cc-noe} Comparison of Thermal NOE and Coupled Cluster in the limit of zero temperature}
There is substantial room for confusion in regards to the extreme values of $f, \bar{f} = 0, 1$ and the zero temperature limit of the theory for the Fermionic problem. Let us try to lay our finger on the critical issues. To make the comparison we have to first allow for the fact that the contractions can become orbital specific and if we introduce factors $f_p, \bar{f}_p = 1 - f_p$, 
the detailed one-electron equations read, in terms of s-amplitudes denoted as $\hat{S}=\sum_{pq} s^p_q \{ \hat{p}^{\dagger} \hat{q} \}$ here
\begin{IEEEeqnarray}{rLl}
-f_q \bar{f}_p \frac{ds^p_{q}}{d \beta} &= f_q \bar{f}_p \left( h^p_{q} + \bar{f}_r \sum_r h^p_{r} s^r_{q} - f_r \sum_r s^p_{r} h^r_{q} - f_r \bar{f}_s \sum_{rs} s^p_{r} h^r_{s} s^s_{q} \right) \nonumber \\
& \qquad -  \mu f_q \bar{f}_p \left( \delta^p_{q} + \bar{f}_p s^p_{q} - f_q s^p_{q}  - f_r \bar{f}_r \sum_r  s^p_{r} s^r_{q} \right) \\
-\frac{ds_0}{d\beta} &= h_0 +  \sum_{p,q} f_p\bar{f}_q h^p_{q} s^q_{p}(\beta) - \mu n_{el}
\end{IEEEeqnarray}
As before we make the substitution $f_q \bar{f}_p s^p_q \rightarrow t^p_q$ and we obtain 
\begin{IEEEeqnarray}{rLl}
 \frac{dt^p_{q}}{d \beta} &= f_q \bar{f}_p  h^p_{q} + \bar{f}_p \sum_r h^p_{r} t^r_{q} - f_q \sum_r t^p_{r} h^r_{q} -  \sum_{rs} t^p_{r} h^r_{s} t^s_{q} \nonumber \\
& \qquad -  \mu \left(  f_q \bar{f}_p \delta^p_{q} + \bar{f}_p t^p_{q} - f_q t^p_{q}  - \sum_r  t^p_{r} t^r_{q} \right) \\
-\frac{ds_0}{d\beta} &= h_0 +  \sum_{p,q}  h^p_{q} t^q_{p}(\beta) - \mu n_{el}
\end{IEEEeqnarray}

To arrive at Coupled Cluster at zero temperature, we now partition the orbitals into occupied $i,j,k$, having $f_i=1, \bar{f}_i=0$ and virtual, $a,b,c$ having $f_a=0, \bar{f}_a=1$. We are not using subtle reasoning that certain $s$-amplitudes might go to $\infty$ at $T=0$, but simply argue that with this choice of $f_p$ the only surviving amplitudes are excitation operators, $\hat{T}=\sum_{a,i} t^a_i \hat{a}^{\dagger} \hat{i}$ and replacing indices accordingly, it is easy to show that the only surviving terms are as follows
\begin{IEEEeqnarray}{rLl}
 \frac{dt^a_{i}}{d \beta} &= \left( h^a_{i} +  \sum_c h^a_{c} t^c_{i} -  \sum_k t^a_{k} h^k_{j} -  \sum_{k,c} t^a_{k} h^k_{c} t^c_{i} \right) \nonumber \\
& \qquad -  \mu  \left(  t^a_{i} - t^a_{i}  \right)  \\  
-\frac{ds_0}{d\beta} &= h_0 +  \sum_{a,i}  h^i_{a} t^a_{i}(\beta) - \mu n_{el}
\end{IEEEeqnarray}
Not surprisingly the chemical potential does not play a role since $\hat{T}$ only excites and preserves the number of electrons. Setting the derivative equal to zero, these equations are exactly the single reference CC equations that would yield the correct ground state.  It is easy to analyse the thermal energy in the case that the hamiltonian is (block)-diagonal such that $h^i_a=0$. We would find $-\frac{ds_0}{d\beta} = h_0$, which would be the ground state energy in that case and this is clearly incorrect. It follows this theory is only applicable at zero temperature, and there it agrees with the thermal NOE equation for the energy. Even at zero T the density matrices are completely different (to such an extent that one does not have density matrices in CC, only excitation amplitudes).  

We can make some pertinent observations to address confusing aspects in the literature (speaking in general terms).
\begin{enumerate}
\item It is perfectly valid to integrate the equations starting from finite (non-extremal) $f_p$ and adjusting the $\beta=0$ limit to have equal populations $n_p = f_p + t^p_p =\frac{n_{el}}{M}$. These equations would yield the correct (Hermitean) reduced density matrix $d^p_q = f^p \delta_{pq} + t^p_q$ that are independent of $f_p$, even as one would \emph{approach} extremal values.
\item Setting the contraction factors $f_p$ to $1$ for some nominally occupied orbitals and to zero for nominally virtual orbitals completely changes the character of the equations. It is impossible to extract the thermal density matrices from the excitation coefficients $t^a_i$. One would have to solve some response equations in addition.
\item At zero temperature both approaches yield correct results, but the solutions for the $t-amplitudes$ are different.
\item One could develop a perturbation theory for one-electron problems using either approach as a starting point. I think these perturbation theories would be different, although the ground state energy would be correct for either approach, and the resulting energies should be the same. 
\item The discussion is essentially the same for interacting systems. The thermal NOE approach is expected to yield different limits in the zero temperature limit than single reference Coupled Cluster Theory (both approximations). In fact we know from our unpublished work that the zero temperature limit from the thermal NOE approach is the connected cumulant approach to the contracted Schr{"\o}dinger equation discussed almost two decades ago \cite{nooijen2003cumulant}. Unfortunately, this approach was abandoned as it appears to have major issues with N-representability. Initial investigations in our group seem to indicate that thermal NOE has similar issues. Let us also mention that thermal NOE appears to be very similar to the approaches suggested by White and Chan \cite{white2018time-dependent} or Mukherjee and coworkers \cite{mandal1998thermal,mandal2001non,mandal2003finite}, except for numerical convergence strategies, and (perhaps) the use of $\beta$-dependent contractions. Many of the subtleties of these methods can be studied using one-electron Hamiltonians for which they are potentially exact. It would be more interesting if they are not exact in that limit perhaps.
\end{enumerate}

\section{\label{sec:conclusion} Concluding remarks and outlook} 
We have come a long way using tools like normal ordering and Wick's theorem that are easy to use, but conceptually very hard, or, one might say, deep. In the few instances that people teach normal ordering in the Chemistry curriculum it is usually in the context of single reference Coupled Cluster theory, and the normal ordering is simply a resorting of annihilation and construction operators, using the (anti)-commutation rules such that so-called quasi-particle annihilation operators are to the right (annihilating the ket $\ket{0}$, while quasi-particle creation operators are to the left, such that they yield 0 when acting on the bra $\bra{0}$. 
As a result $\bra{0} \left \lbrace \Omega_{\lambda} \right \rbrace \ket{0} =0$ for any non-empty string of operators $\Omega_{\lambda}$ or linear combinations. Upon the introduction (seldom proved) of Wick's theorem we know then that only \emph{fully connected} terms contribute. In the context of this paper \emph{normal ordering} is a misnomer. I
t is not possible to write the operator $\left \lbrace \hat{p}^{\dagger} \hat{q} \right \rbrace$ as a particular order of $\hat{p}^{\dagger}$ and $\hat{q}$. 
The only way to think about it is as $\left \lbrace \hat{p}^{\dagger} \hat{q} \right \rbrace = \hat{p}^{\dagger} \hat{q} - C_{pq}$, where $C_{pq}$ is a constant (or sometimes a function), callled the contraction of the operators. 
It is not hard to understand the definition, but it can be hard to really understand the power of such a simple devise. 

Starting from the most traditional quantum chemistry approaches using Slater rules or elementary second quantization it appears impossible to tackle the problems of Statistical Mechanics that involve taking the trace over all determinants (in Fock space!). That is Full CI on steroids. Using the mechanism of normal ordering, as we advertise here, one defines a constant zero order density matrix $D_0$ (a multiple of the unit matrix in Fock space) 
and defines the contraction as $\hat{p}^{\dagger} \cdot \hat{q} = Tr (\hat{p}^{\dagger} \hat{q} D_0)$ 
and as a result  $Tr (\left \lbrace \hat{p}^{\dagger} \hat{q} \right \rbrace  D_0) = 0$. Wick's theorem does not depend in any way on the initial poetry to set things up. It is not physics but rather a mathematical theorem that works on algebraic structures as we are employing here.

The other ingredient is the use of the normal ordered exponential to parameterize the true density matrix $\hat{D} = \left \lbrace e^{\hat{S} }\right \rbrace \hat{D}_0$. Could we not have used an ordinary exponential? Yes, in fact we can! $\hat{D} = \frac{1}{Tr(D_0)} e^{- \beta \hat{H}} \hat{D}_0$. Unfortunately that is just a restatement of the problem. The normal ordering is key to the solution. Most of the other manipulations in this paper are merely technical, in particular the convenient reduction of the propagation equations to a connected form (see appendix).
\begin{eqnarray}
\braket { \{ \hat{\Omega}_{\lambda} \} \frac{d \hat{S}}{d \tau} } = \braket { \{ \hat{\Omega}_{\lambda} \} \big[ \hat{H} \left \lbrace e^{\hat{S}} \right \rbrace \big]_{connected} }
\end{eqnarray}

We spend quite some time on the fact that the choice of the contraction $f$ is essentially arbitrary. The $s$ or $t$-amplitudes adjust such that the reduced density matrix $d^p_q(\tau) = f \delta_{pq} + t^p_q(\tau)$ are invariant under the choice of $f$. With a little extra work one can in fact remove all singles $t^p_q$ amplitudes and factors $f$ and use $d^p_q$ and $\bar{d}^p_q = \bar{f} - t^p_q = \delta_{pq} - t^p_q$. This would constitute a proof that the results do not depend on the choice of $f$ and in addition this would have allowed us to make a clear connection to the connected cumulant formulation \cite{nooijen2003cumulant} at zero temperature. We also emphasized here the pitfalls one can fall into if one makes the choice $f=0$, or rather, if one makes this choice too quickly.

In this paper we discussed a number of illustrative examples that all yield the exact (numerical) result provided the differential equations are initialized properly. For Fermionic one-electron problems one obtains the usual results from Fermi-Dirac theory, initializing the theory at the high T limit or $\beta=0$. For harmonic oscillators the high T-limit is less appropriate, while also T=0 is hard because the inital propagation is essentially flat (all derivatives are zero). We started the propagation by extracting the $t$-amplitudes from density matrices obtained from a small sum over state calculation at low temperature. Another interesting illustration is the calculation of harmonic Franck-Condon spectra, where we also show that we do not need to use the normal-ordered exponential, but it is more convenient to use the usual exponential and (commuting) excitation operators only (using f=0).

These exactly solvable problems are useful to illustrate the validity of the theory but other problems are truely of interest. The most obvious candidate is the many-body electronic structure case. There is one glaring problem and this is that the residual equations are exactly the same (or can be interconverted by simple substitutions) as connected cumulant theory \cite{nooijen2003cumulant}. This theory was a major dissapointment: it took us two years to publish it (i.e. to write the paper) after the work was completed. Moreover, this paper essentially killed the field as more recent approaches there are more wave function like in spirit (in spite of terminology). In the present thermal context we spent another 3 months on its implementation and application to strongly correlated systems with similar dissapointments and clear violations of N-representability. That was also two years ago. The theory as exposed here is too beautiful not to publish. Better ideas are needed to push it towards application for relevant electronic structure problems. We note that it may work for metallic systems in the current form. The multireference nature of metals is not sufficiently clear to these authors.

Another application is to vibrational and also non-adiabatic vibronic problems. We think the application to (anharmonic) vibrational problems is in principle straightforward. In our group we are working on the application of both time-autocorrelation functions and thermal properties for non-adiabatic vibronic problems. The theory is less straightforward here, in particular we did not find a connected form of the equations and this remains work in progress.

A very appealing application would be to electron-phonon models in a solid state context. The theory is tailor made for these problems using both bosonic and Fermionic second quantization. We think the main issue is the same as the problems plaguing the pure electronic theory: N-representability.

Let us finally address a glaring ommission in this paper. A lot of work is ongoing in the recent  quantum chemistry literature that tackles the same problems as we do here. There are clearly close connections in regards to the use of normal ordered exponentials and the use of Wick's theorem. As far as we are aware the other approaches use time- or temperature-dependent contractions, starting from the interaction picture. That requires some changes and it is not completely straightforward to connect these theories (e.g. \cite{mandal1998thermal, mandal2001non, mandal2003finite, hermes2015finite,white2018time-dependent}. We know for sure the zero-temperature limit of our electronic structure formulation is the connected contracted Schr{"\o}dinger equation and not single reference coupled cluster theory. Due to the 'vagueries' of the temperature dependent contractions we are not sure at present about other formulations. We would caution care in claiming the connection to Coupled Cluster, referring to our analysis in \Sref{sec:reflections}. We tried hard here to have this paper be self-contained. The only assumptions are second quantization and Wick's theorem. All the rest can be (and is essentially) derived in this paper. This suffices for now and we forgo detailed comparisons or reflections on the extant literature.

\section*{Acknowledgements} 
This research has been supported by the Natural Sciences and Engineering Research Council of Canada (NSERC). It is our pleasure to contribute to the festschrift for Prof. John F. Stanton. MN has had many enjoyable and fruitful interactions with JFS in the past and we think it is fitting to contribute a pedagogically oriented paper to commemorate the joyful occasion of John's 60th birthday.

\begin{appendices}
\section{ \label{sec:wick} Reduction of amplitude equations to connected form}
One of the important features that greatly simplify the CC amplitude equations is the reduction of the equation to a connected form in which each amplitude is contracted to the Hamiltonian. In this appendix we prove the crucial relation
\begin{equation}
\hat{H} \left\lbrace e^{\hat{S}(\beta)}\right\rbrace =\left\lbrace \big[ \hat{H} \left\lbrace e^{\hat{S}(\beta)} \right\rbrace \big]_{connected} e^{\hat{S}(\beta)} \right\rbrace
\end{equation}

For the left hand side, expand the exponential ansatz into Taylor series
\begin{equation}
\hat{H} \left\lbrace e^{\hat{S}(\beta)} \right\rbrace =\sum_n \frac{1}{n!}\hat{H} \left\lbrace  \hat{S}(\beta)^n \right\rbrace
\end{equation}
Applying Wick's Theorem, each term in the Taylor series can be expanded and we can distinguish $\hat{S}$ amplitudes that are connected to the Hamiltonian and those that are not. If in a particular term of power $n$, $k$ amplitudes $\hat{S}$ are connected to $\hat{H}$, indicated by an overbar, while $n-k$ are not connected to $\hat{H}$, the number of distinct such contributions is given by 
\begin{equation}
\label{eqn:proof_2}
\binom{n}{k}=\frac{n!}{k!(n-k)!}
\end{equation}
and one can write
\begin{eqnarray}
\label{eqn:proof_1}
\sum_n \frac{1}{n!} \hat{H} \left\lbrace \hat{S}(\beta)^n \right\rbrace &= \sum_n 
  \sum_{k=0}^n  \binom{n}{k} \left\lbrace (\overline{ \hat{H} \left\lbrace \hat{S}(\beta)^k\right\rbrace } (\hat{S}(\beta)^{n-k}) \right\rbrace \nonumber \\
  &= \sum_m 
  \sum_k  \binom{m+k}{k} \left\lbrace (\overline{ \hat{H} \left\lbrace \hat{S}(\beta)^k\right\rbrace } (\hat{S}(\beta)^{m}) \right\rbrace 
  \end{eqnarray}
From Eq(\ref{eqn:proof_1}) and Eq(\ref{eqn:proof_2}) we get the desired result:
\begin{eqnarray}
\hat{H} \left\lbrace e^{\hat{S}(\beta)} \right\rbrace  &= \sum_k \frac{1}{k!} \left\lbrace  \overline { \hat{H} \left \lbrace \hat{S}(\beta)^k \right\rbrace } \sum_m (\frac{1}{(m)!} \hat{S}(\beta)^{m}) \right\rbrace  \nonumber \\
&=\left\lbrace \big[ \hat{H} \left\lbrace e^{\hat{S}(\beta)}  \right\rbrace \big]_{connected} e^{\hat{S}(\beta)} \right\rbrace
\end{eqnarray}
It may be instructive to do the same proof for an ordinary exponential of commuting operators, as occurs for example in single reference coupled Cluster theory
\begin{eqnarray}
\hat{H}e^{\hat{T}} &= e^{\hat{T}} \big( e^{-\hat{T}}  \hat{H} e^{\hat{T} } \big) \nonumber \\
&= e^{\hat{T}} \big( \hat{H} e^{\hat{T}} \big)_{connected} \nonumber \\
&= \left\lbrace \big( \hat{H} e^{\hat{T}} \big)_{connected} e^{\hat{T}} \right \rbrace
\end{eqnarray}
The second line is then usually proven by applying the Baker-Campbell-Hausdorf expansion to obtain a nested commutator, and showing that each commutator implies only terms survive in which $\hat{T}$ is contracted to the Hamiltonian. The above proof based on Wick's theorem is in some sense easier and highlights the role of the $\frac{1}{n!}$ coefficients in combination with the binomial "choose" factors.

\end{appendices}

\bibliographystyle{jcp}
\bibliography{citations}

\begin{thebibliography}{10}

\bibitem{fetter2012quantum}
{\sc A.~L. Fetter} and {\sc J.~D. Walecka},
\newblock {\em Quantum theory of many-particle systems},
\newblock Courier Corporation, 2012.

\bibitem{abrikosov2012methods}
{\sc A.~A. Abrikosov}, {\sc L.~P. Gorkov}, and {\sc I.~E. Dzyaloshinski},
\newblock {\em Methods of quantum field theory in statistical physics},
\newblock Courier Corporation, 2012.

\bibitem{negele2018quantum}
{\sc J.~W. Negele},
\newblock {\em Quantum many-particle systems},
\newblock CRC Press, 2018.

\bibitem{mattuck1992guide}
{\sc R.~D. Mattuck},
\newblock {\em A guide to Feynman diagrams in the many-body problem},
\newblock Courier Corporation, 1992.

\bibitem{matsubara1955new}
{\sc T.~Matsubara},
\newblock {\em Progress of theoretical physics} {\bf 14}, 351 (1955).

\bibitem{bloch1958developpement}
{\sc C.~Bloch} and {\sc C.~De~Dominicis},
\newblock {\em Nuclear Physics} {\bf 7}, 459 (1958).

\bibitem{hirata2013kohn}
{\sc S.~Hirata} and {\sc X.~He},
\newblock {\em The Journal of chemical physics} {\bf 138}, 204112 (2013).

\bibitem{hirata2013second}
{\sc S.~Hirata}, {\sc X.~He}, {\sc M.~R. Hermes}, and {\sc S.~Y. Willow},
\newblock {\em The Journal of Physical Chemistry A} {\bf 118}, 655 (2013).

\bibitem{he2014finite}
{\sc X.~He}, {\sc S.~Ryu}, and {\sc S.~Hirata},
\newblock {\em The Journal of Chemical Physics} {\bf 140}, 024702 (2014).

\bibitem{santra2017finite}
{\sc R.~Santra} and {\sc J.~Schirmer},
\newblock {\em Chemical Physics} {\bf 482}, 355 (2017).

\bibitem{welden2016exploring}
{\sc A.~R. Welden}, {\sc A.~A. Rusakov}, and {\sc D.~Zgid},
\newblock {\em The Journal of chemical physics} {\bf 145}, 204106 (2016).

\bibitem{kananenka2016efficient}
{\sc A.~A. Kananenka}, {\sc J.~J. Phillips}, and {\sc D.~Zgid},
\newblock {\em Journal of chemical theory and computation} {\bf 12}, 564
  (2016).

\bibitem{kananenka2016efficientb}
{\sc A.~A. Kananenka}, {\sc A.~R. Welden}, {\sc T.~N. Lan}, {\sc E.~Gull}, and
  {\sc D.~Zgid},
\newblock {\em Journal of chemical theory and computation} {\bf 12}, 2250
  (2016).

\bibitem{zgid2017finite}
{\sc D.~Zgid} and {\sc E.~Gull},
\newblock {\em New Journal of Physics} {\bf 19}, 023047 (2017).

\bibitem{sanyal1992thermal}
{\sc G.~Sanyal}, {\sc S.~H. Mandal}, and {\sc D.~Mukherjee},
\newblock {\em Chemical physics letters} {\bf 192}, 55 (1992).

\bibitem{mandal1998thermal}
{\sc S.~H. Mandal}, {\sc G.~Sanyal}, and {\sc D.~Mukherjee},
\newblock A thermal cluster-cumulant theory,
\newblock in {\em Microscopic Quantum Many-Body Theories and Their
  Applications}, pp. 93--117, Springer, 1998.

\bibitem{mandal2001non}
{\sc S.~H. Mandal}, {\sc R.~Ghosh}, and {\sc D.~Mukherjee},
\newblock {\em Chemical physics letters} {\bf 335}, 281 (2001).

\bibitem{mandal2003finite}
{\sc S.~H. Mandal}, {\sc R.~Ghosh}, {\sc G.~Sanyal}, and {\sc D.~Mukherjee},
\newblock {\em International Journal of Modern Physics B} {\bf 17}, 5367
  (2003).

\bibitem{hermes2015finite}
{\sc M.~R. Hermes} and {\sc S.~Hirata},
\newblock {\em The Journal of chemical physics} {\bf 143}, 102818 (2015).

\bibitem{white2018time-dependent}
{\sc A.~F. White} and {\sc G.~K.-L. Chan},
\newblock {\em Journal of Chemical Theory and Computation} {\bf x}, xx (2018).

\bibitem{hummel2018finite}
{\sc F.~Hummel},
\newblock {\em Journal of Chemical Theory and Computation}  (2018).

\bibitem{nooijen1992thesis}
{\sc M.~Nooijen},
\newblock The coupled cluster Green’s function, 1992.

\bibitem{nooijen1992coupled}
{\sc M.~Nooijen} and {\sc J.~G. Snijders},
\newblock {\em International Journal of Quantum Chemistry} {\bf 44}, 55 (1992).

\bibitem{nooijen1993coupled}
{\sc M.~Nooijen} and {\sc J.~G. Snijders},
\newblock {\em International journal of quantum chemistry} {\bf 48}, 15 (1993).

\bibitem{monkhorst1981recursive}
{\sc H.~J. Monkhorst}, {\sc B.~Jeziorski}, and {\sc F.~E. Harris},
\newblock {\em Physical Review A} {\bf 23}, 1639 (1981).

\bibitem{harris1992algebraic}
{\sc F.~E. Harris}, {\sc H.~Monkhorst}, and {\sc D.~L. Freeman},
\newblock (1992).

\bibitem{shavitt2009many}
{\sc I.~Shavitt} and {\sc R.~J. Bartlett},
\newblock {\em Many-body methods in chemistry and physics: MBPT and
  coupled-cluster theory},
\newblock Cambridge university press, 2009.

\bibitem{nooijen2003cumulant}
{\sc M.~Nooijen}, {\sc M.~Wladyslawski}, and {\sc A.~Hazra},
\newblock {\em The Journal of chemical physics} {\bf 118}, 4832 (2003).

\bibitem{nooijen1996general}
{\sc M.~Nooijen} and {\sc R.~J. Bartlett},
\newblock {\em The Journal of chemical physics} {\bf 104}, 2652 (1996).

\bibitem{nooijen1999similarity}
{\sc M.~Nooijen},
\newblock {\em Spectrochimica Acta Part A: Molecular and Biomolecular
  Spectroscopy} {\bf 55}, 539 (1999).

\bibitem{faucheaux2015higher}
{\sc J.~A. Faucheaux} and {\sc S.~Hirata},
\newblock {\em The Journal of chemical physics} {\bf 143}, 134105 (2015).

\bibitem{faucheaux2017diagrammatic}
{\sc J.~A. Faucheaux},
\newblock {\em Diagrammatic theories for the vibrational many-body problem},
\newblock PhD thesis, University of Illinois at Urbana-Champaign, 2017.

\bibitem{faucheaux2018similarity}
{\sc J.~A. Faucheaux}, {\sc M.~Nooijen}, and {\sc S.~Hirata},
\newblock {\em The Journal of chemical physics} {\bf 148}, 054104 (2018).

\end{thebibliography}

\end{document}